\documentstyle[twoside,12pt,epsf]{article}
\let\ssection=\section
\renewcommand{\section}{\setcounter{equation}{0}\ssection}

\def\head#1#2{
\markboth{}{}
\setcounter{page}{#1}
\setcounter{section}{0}
\begin{huge}
\begin{flushleft}
\noindent\hangindent\parindent
{#2}
\end{flushleft}

\end{huge}
\bigskip
}

\def\auaddr#1#2{
{\noindent\LARGE\em#1}
{\noindent #2}
\medskip}

\def\and{
{\LARGE\em \&}
\bigskip
}

\def\refce{\smallskip \noindent\hangindent\parindent}

\def\caption#1{{ \begin{quote} \rm #1 \end{quote} }}

\newskip\humongous \humongous=0pt plus 1000pt minus 1000pt
\def\caja{\mathsurround=0pt}
\def\eqalign#1{\,\vcenter{\openup1\jot \caja
        \ialign{\strut \hfil$\displaystyle{##}$&$
        \displaystyle{{}##}$\hfil\crcr#1\crcr}}\,}
\newif\ifdtup

\font\ensemble=msbm10

\def\RR{\hbox{\ensemble R}}
\def\ZZ{\hbox{\ensemble Z}}

\def\CC{\hbox{\ensemble C}}

\begin{document}


\thispagestyle{empty}

\begin{center}

\hfill CPHT/CL-615-0698\\
\hfill hep-th/9806199\\

\vskip .6in

{\large \bf  Lectures on D-branes}
\vskip .4in

{\bf Constantin P.  Bachas}\footnote{Address after Sept. 1:
  Laboratoire de Physique Th{\'e}orique, Ecole Normale Sup{\'e}rieure, 
 24  rue Lhomond, 75231 Paris, FRANCE, \em 
email~: bachas@physique.ens.fr }

\vskip .2in

{ Centre de Physique Th{\'e}orique, 
Ecole Polytechnique \\
 91128 Palaiseau, FRANCE\\
 \vskip 0.2cm
\em bachas@cpht.polytechnique.fr}

\vskip .15in

\vskip .15in

\end{center}

\vskip .2in

\begin{center} {\bf ABSTRACT } \end{center}
\begin{quotation}\noindent

This is an introduction to the physics of  D-branes. Topics covered
include Polchinski's original calculation, a critical assessment of
some duality checks, D-brane scattering, and effective worldvolume actions.
Based on lectures  given in  1997 at the Isaac Newton Institute,
Cambridge, at the Trieste Spring School on String Theory, and at the 
31rst International Symposium Ahrenshoop  in Buckow. 

\end{quotation}
\vskip 1.0cm
\vskip 2.cm
June  1998\\

\vfill
\eject


\thispagestyle{plain}
\def\tit{ Lectures  on D-branes}
\def\auth{Constantin Bachas}
\head{1}
{\tit}
\auaddr{\auth}{}
\markboth{ Bachas}{  Lectures  on D-branes}
\def\ve{\varepsilon}

\section{Foreword}

   Referring in his `Republic' to stereography -- the study of solid
 forms  -- Plato was saying~: {\it  ... for even now, neglected and
   curtailed as it is, not only by the many but even by professed
   students, who can suggest no use for it, nevertheless in the face
   of all these obstacles it makes progress on account of its
   elegance, and it would not be astonishing if it were unravelled.}
\footnote{Translated  by Ivor Thomas in `Greek Mathematical Works',
   Loeb Classical Library, Harvard U. Press 1939.} 
Two and a half millenia later, much of this could have been said for
 string theory.  The subject has
progressed over the years by leaps and bounds, despite periods of
 neglect and  (understandable) criticism for  lack of 
direct experimental input. To be sure, the construction and  
 key ingredients of the theory --
gravity, gauge invariance, chirality -- have a  firm empirical
 basis, yet what has often  catalyzed progress is
 the  power and elegance of the
underlying ideas,  which look (at least a posteriori) inevitable. 
 And  whether the ultimate  structure will be 
unravelled or not, there
is already a  name waiting for `it': ${\cal M}$ theory.

  Few of the features of the theory,
 uncovered so far,  exemplify this power and
elegance better than D-branes.  Their definition   as
allowed endpoints for  open strings, generalizes    the
notion of quarks on which the QCD string can terminate.
In contrast to the quarks of QCD, D-branes are however  intrinsic 
excitations of the fundamental  theory: their  existence is
required   for consistency, and  their
properties -- mass, charges, dynamics -- are unambiguously
determined in terms of the Regge slope  $\alpha^\prime$ and 
the asymptotic values of the dynamical  moduli.
They resemble  in these respects conventional
field-theory  solitons, from which however they differ in important
ways.  D-particles, for instance,  can probe distances much smaller than
the size of the fundamental-string quanta,  at weak coupling. 
In any case, D-branes, fundamental strings and smooth solitons fill
together  the multiplets of the various (conjectured) dualities, which
connect all string theories to each other. D-branes  have, in this sense, 
played a crucial role  in delivering  the  important message of the 
`second string revolution',  that the way to  
reconcile quantum mechanics and Einstein gravity may be so constrained
as to  be `unique'.

   Besides filling duality multiplets, D-branes have however also
opened a window into the microscopic structure of quantum
gravity. The D-brane model of black  holes may prove as important for
understanding black-hole thermodynamics, as has  the Ising model
proven in the past  for understanding second-order   phase transitions.
Technically,  the D-brane concept is so powerfull because of
the surprising  relations it has revealed
between supersymmetric gauge theories and  geometry.
These relations   follow from the fact that 
Riemann surfaces with boundaries admit dual interpretations  as  field-theory
diagrams along various  open- or closed-string channels. 
Thus, in particular,  the counting of microscopic BPS states of a black hole,
an ultraviolet problem of quantum gravity, can be mapped to 
the more familiar problem of studying the moduli space of supersymmetric
gauge theories. Conversely, `brane engineering' has been a 
 useful  tool  for discussing  Seiberg dualities and other infrared
properties of supersymmetric gauge theories, while 
low-energy  supergravity, corrected by
classical string effects, may offer a new line of attack on the old
problem of solving
gauge theories in the planar ('t Hooft)  limit.

   Most of these exciting developments will  not be discussed in the
present lectures. The material included here covers only some of the
earlier papers on D-branes, and is a modest expansion of a previous
`half lecture' by the author (Bachas 1997a). The main difference  from
other existing reviews of the same subject
 (Polchinski {\em et al} 1996,  Polchinski 1996, Douglas 1996,
Thorlacius 1998, Taylor 1998) is  in the  
emphasis and presentation style.
The aim is to provide the reader (i) with
a basis, from which to move on  to reviews of  related and/or
more advanced topics,  and (ii)  with an extensive (though far from
complete)  guide to the literature. I will be assuming a working
knowledge of perturbative string theory at the level of Green, Schwarz
and Witten  (1987) 
(see also Ooguri and Yin 1996,  Kiritsis 1997, Dijkgraaf 1997,   
and volume one  of Polchinski 1998,  for recent reviews),
and  some familiarity with the main ideas of
string duality,  for which there exist many  nice and 
complimentary lectures  (Townsend 1996b and  1997,
Aspinwall 1996, Schwarz 1997a and 1997b, Vafa 1997, Dijkgraaf 1997,
F{\"o}rste and Louis 1997, de Wit and Louis 1998, L{\"u}st 1998, 
Julia 1998, West in this volume,
Sen in this volume).  

   A list of pedagogical reviews 
for further reading includes~:
 Bigatti and Susskind (1997), 
Bilal (1997), Banks (1998), Dijkgraaf {\em et al} (1998), 
 and de Wit (1998) for the Matrix-model conjecture, 
Giveon and Kutasov (1998)  for brane engineering of gauge theories,
Maldacena (1996) and Youm (1997) for the D-brane approach to black
holes,  Duff {\em et al} (1995),
Duff (1997), Stelle (1997,1998),  Youm (1997) 
and Gauntlett (1997)  for
reviews of branes  from the complimentary, supergravity viewpoint.
I am not aware of any extensive reviews  of type-I compactifications,
of D-branes in general curved backgrounds, and of
semiclassical calculations using  D-brane instantons. Some short
lectures on these subjects, which the reader may consult for further
references, include Sagnotti (1997), Bianchi (1997), Douglas (1997),
 Green (1997),
Gutperle (1997),  Bachas (1997b), Vanhove (1997) and Antoniadis
 {\em et al} (1998).
Last but not least, dualities  in  rigid
supersymmetric field theories -- a subject intimately tied  to
D-branes -- are  reviewed by Intriligator and Seiberg (1995), 
Harvey (1996), Olive (1996), Bilal (1996),
Alvarez-Gaume and Zamora (1997), 
Lerche (1997), Peskin (1997), Di Vecchia (1998)  and  West (1998).

\section{Ramond-Ramond fields}  

 With the exception of the heterotic string, all other consistent
string theories  contain in their spectrum antisymmetric tensor
fields  coming from the  Ramond-Ramond sector. This is the case for the
type-IIA and type-IIB superstrings, as well as for the type-I theory whose
closed-string states are a subset of those of the type-IIB.
One of the key properties of D-branes is that they are the elementary charges
of Ramond-Ramond fields, so let us  begin  the discussion
by recalling some basic facts about these fields.

\subsection{Chiral bispinors}
\label{subsec:bispinor}

The states  of a closed-string theory are given by 
the tensor product of a left- and a right-moving worldsheet
sector. For type-II theory in the covariant (NSR)  formulation,
each sector  contains at the massless level a
ten-dimensional vector and a ten-dimensional Weyl-Majorana spinor. This is
depicted figuratively as follows:
$$
\Bigl( \vert \mu\rangle  \oplus \vert a \rangle \Bigr)_{left}
 \ \otimes \ 
\Bigl( \vert \nu\rangle  \oplus \vert b \rangle \Bigr)_{right}
\ \ \ , 
$$
where
 $\mu,\nu = 0,... ,9$ and $a,b = 1,... ,16$
 are,
respectively,  vector and   spinor indices.
Bosonic fields thus include a two-index tensor,  which can be
decomposed into  a symmetric traceless, a  trace,  and an antisymmetric
part: these are the usual fluctuations of the  graviton
 ($G_{\mu\nu}$), dilaton ($\Phi$) and Neveu-Schwarz 
antisymmetric tensor ($B_{\mu\nu}$). In addition,  massless bosonic
fields include a Ramond-Ramond bispinor $H_{ab}$,
defined as the polarization in the corresponding
vertex operator
\begin{equation}
V_{\rm RR} \sim  \int d^2\xi  \; e^{ip^\mu X_\mu} \; 
{\overline S}^{\; \rm T} \Gamma^0 H(p) S  \ .
\label{eq:vertex}
\end{equation}
In this expression
  $S^a$ and  ${\overline  S}^{\;b}$ are
 the covariant left- and right-moving fermion emission
operators -- a product of the corresponding spin-field and ghost
  operators (Friedan {\em et al}  1986), 
$p^\mu$ is the ten-dimensional  momentum, 
and $\Gamma^0$ the ten-dimensional gamma matrix.

The  bispinor field $H$  can be decomposed in
a complete basis of all gamma-matrix
antisymmetric products
\begin{equation}
 H_{ab} = \sum_{n=0}^{10}\; {i^{n}\over n!}\;
  H_{\mu_1...\mu_n}
(\Gamma^{\mu_1...\mu_n})_{ab}\ .
\label{eq:decomposition}
\end{equation}
Here
$
 \Gamma^{\mu_1...\mu_n} \equiv {1\over n!}\; \Gamma^{[\mu_1}...
\Gamma^{\mu_n]}$, where 
square brackets denote   the  alternating sum
over all permutations of the enclosed indices,   
 and the $n=0$ term
 stands by convention for the
identity in spinor space.  I use the following conventions~:   
 the ten-dimensional  gamma matrices are purely imaginary
and obey the algebra $\{ \Gamma^\mu, \Gamma^\nu \} = -2\eta^{\mu\nu}$
with metric signature $(-+...+)$.
The chirality operator is
 $\Gamma_{11} = \Gamma^0\Gamma^1...\Gamma^9$,
Majorana spinors are real,  and the Levi-Civita
tensor $\epsilon^{01...9}= 1$. 

 In view of the  decomposition (\ref{eq:decomposition}),  the Ramond-Ramond
 massless fields are a collection of antisymmetric
Lorentz tensors.
These tensors  are not  independent because 
 the bispinor  field is subject to  definite chirality projections, 
\begin{equation}
 H  =  \Gamma_{11}\; H =  \pm  H\;  \Gamma_{11}  \ .
\label{eq:chiral}
\end{equation}
The choice of sign distinguishes between the type-IIA and
type-IIB models. For the type-IIA theory $S$ and $\overline S$ have  opposite
chirality, so one should choose the sign plus. In the type-IIB case,
on the other hand, 
the two spinors have the same chirality and one should 
choose the sign minus. 
To express  the chirality constraints 
in terms of the antisymmetric  tensor fields we use
the gamma-matrix  identities
\begin{equation}
\Gamma^{\mu_1...\mu_n} \Gamma_{11} =
(-)^n  \Gamma_{11} \Gamma^{\mu_1...\mu_n} = 
 { \epsilon^{\mu_1...\mu_{10}}  \over (10-n)!} \;
 \Gamma_{\mu_{10}...\mu_{n+1}}
\end{equation}
 It follows easily
that only even-$n$ (odd-$n$) terms are allowed in the type-IIA
(type-IIB) case. Furthermore the  antisymmetric tensors
obey the  duality relations
\begin{equation}
H^{\mu_1...\mu_n} = { \epsilon^{\mu_1...\mu_{10}}  \over (10-n)!} \;
 H_{\mu_{10}...\mu_{n+1}} \ ,  \ \ \ 
{\rm or}\ \  {\rm equivalently} \ \ \  H^{(n)}= \ ^*H^{(10-n)}\ .   
\label{eq:dual}
\end{equation}
 As a check note that the type-IIA  theory
has  independent tensors with $n=0,2$ and $4$ indices, while
the type-IIB  theory has $n=1,3$ and a self-dual $n=5$ tensor.
The number of independent tensor components adds up in both
cases to $16\times 16 = 256$:
\begin{eqnarray}
&{ {\rm IIA:}}&\ 
1 + {10\times 9 \over 2!} + {10\times 9 \times 8 \times 7 \over
 4!}\;
 =\; 256\ , \nonumber \\
& & \nonumber \\
&{ {\rm IIB:}}& \  
10 + {10\times 9 \times 8 \over 3!} + {10\times 9 \times 8
\times 7 \times 6 \over 2 \times  5!}\; =\; 256\ .  \nonumber
\end{eqnarray}
This  is  precisely  the number of components of a bispinor.

  Finally let us consider the  type-I theory, which can be thought of
  as an orientifold projection of type-IIB  (Sagnotti
  1988, Ho\v rava 1989a). This projection involves an interchange of left- and
  right-movers on the worldsheet. The surviving closed-string states
  must be symmetric in the Neveu-Schwarz sector and antisymmetric in
  the Ramond-Ramond sector, consistently with supersymmetry and with
  the fact that the graviton should survive. This implies the extra
  condition on the bispinor field
\begin{equation}
  (\Gamma^0 H)^{\rm T} = - \Gamma^0 H \ .
\end{equation} 
Using  $(\Gamma^\mu)^{\rm T} = - \Gamma^0 \Gamma^\mu \Gamma^0$ we
conclude, after some straightforward algebra,  
that the only Ramond-Ramond fields  surviving the  extra projection are
$H^{(3)}$  and its dual,  $H^{(7)}$.

\subsection{Supergraviton multiplets}
\label{subsec:freeeqs}

  The mass-shell or super-Virasoro
conditions for the vertex operator $V_{RR}$
imply that the bispinor field obeys two
 massless Dirac equations, 
\begin{equation}
 \slash \hskip -0.2cm  { p}  H = H \slash \hskip -0.2cm  { p}
 = 0 \ .
\end{equation}
To convert these to equations for the tensors we need  the gamma
identities
\begin{equation}
\eqalign{
\Gamma^\mu \Gamma^{\nu_1...\nu_n}& =  \Gamma^{\mu\nu_1...\nu_n}
- {1\over (n-1)!}\; \eta^{\mu [\nu_1}\; \Gamma^{\nu_2...\nu_n]}
\cr
 \Gamma^{\nu_1...\nu_n} \Gamma^\mu & =  \Gamma^{ \nu_1...\nu_n\mu}
- {1\over (n-1)!}\; \eta^{\mu [\nu_n}\; \Gamma^{\nu_1...\nu_{n-1}]} \cr}
\end{equation}
and the decomposition (\ref{eq:decomposition}) of a bispinor. 
 After some
straightforward algebra one finds
\begin{equation}
 p^{[\mu} H^{\nu_1...\nu_n]} = p_\mu H^{\mu \nu_2...\nu_n} =
 0 \ .
\end{equation}
These are  the Bianchi identity and free massless equation for
an antisymmetric tensor field strength in momentum space, which we may
write in more economic form as 
\begin{equation}
 d H^{(n)} = d \ ^*H^{(n)} = 0
\end{equation} 
The polarizations of covariant Ramond-Ramond emission vertices
are therefore field-strength tensors  rather than gauge potentials.

Solving the Bianchi identity locally
allows us to express the  $n$-form  field strength as the
exterior derivative of a $(n-1)$-form potential
\begin{equation}
H_{\mu_1...\mu_n} =
{1\over (n-1)!}\; \partial_{\ [ \mu_1} C_{\mu_2...\mu_n]}  ,
\ \ {\rm or}\ \ \  H^{(n)} = d C^{(n-1)} \ .
\end{equation}
Thus the type-IIA theory has a vector ($C_\mu$) and a three-index
tensor potential ($C_{\mu\nu\rho}$) , in addition to a constant
non-propagating zero-form field strength ($H^{(0)}$), while the
type-IIB theory has a zero-form ($C$), a two-form ($C_{\mu\nu}$)
and a four-form potential ($C_{\mu\nu\rho\sigma}$), the latter
 with self-dual field strength. Only the two-form potential survives
 the type-I orientifold projection. These facts are summarized in 
 table 1. A $(p+1)$-form `electric' potential can of course be
 traded for a $(7-p)$-form `magnetic' potential, obtained by solving
 the Bianchi identity of the dual field strength. 
\vskip 0.3cm

\begin{table}[htp]
\begin{center}
\begin{tabular}{|c|c|c|}
\cline{2-3}
\multicolumn{1}{l|}{} & \multicolumn{1}{c|}{ }& \multicolumn{1}{c|}{ } \\
\multicolumn{1}{l|}{} & \multicolumn{1}{c|}{Neveu-Schwarz}
& \multicolumn{1}{c|}{ Ramond-Ramond} \\
\multicolumn{1}{l|}{} & \multicolumn{1}{c|}{ }& \multicolumn{1}{c|}{ } \\
 \hline
 & & \\
type-IIA & $G_{\mu\nu}, \Phi , B_{\mu\nu}$ &  $C_\mu , C_{\mu\nu\rho}$
~; 
$H^{(0)}$  \\
 & & \\
 \hline
 & &  \\
type-IIB & $G_{\mu\nu}, \Phi , B_{\mu\nu}$ &  $C, C_{\mu\nu} ,
 C_{\mu\nu\rho\tau}$
 \\
 & & \\ 
 \hline
 & & \\
type-I  & $G_{\mu\nu}, \Phi $ &  $ C_{\mu\nu}$
\\
& & \\
  \hline
 & & \\ 
heterotic   & $G_{\mu\nu}, \Phi , B_{\mu\nu} $ & \\ 
& & \\ 
 \hline
\end{tabular}
\end{center}
\caption{
String origin of massless fields completing the N=1 or  N=2
supergraviton multiplet of the various theories in ten dimensions.}
\end{table}

From the point of view of  low-energy supergravity all Ramond-Ramond
fields belong to  the ten-dimensional graviton
multiplet. For N=2 supersymmetry this contains 128  bosonic helicity
states, while for N=1 supersymmetry it only contains 64. For both the
type-IIA and type-IIB theories, half of these states come from the
Ramond-Ramond sector~, as can be checked by counting the transverse
physical components of the gauge potentials~:
\begin{eqnarray}
& { {\rm IIA:}}& \ \ \
8 + {8\times 7\times 6 \over 3!}= 64\ , \nonumber
\\& & \nonumber  \\
& { {\rm IIB:}}& \ \ \
1 + {8\times 7  \over 2!} + {8\times 7 \times 6
\times 5   \over 2 \times  4!} = 64\ .  \nonumber
\end{eqnarray}
This  counting is simpler in the light-cone Green-Schwarz formulation,
where the Ramond-Ramond fields correspond to a  chiral SO(8) bispinor.

\subsection{Dualities and   RR  charges}
\label{subsec:nocharge}

A $(p+1)$-form potential couples naturally to a $p$-brane,
i.e. an excitation extending over  $p$ spatial dimensions.
Let $Y^\mu(\zeta^\alpha)$ be the worldvolume  of the
brane ($\alpha =0,...,p$), and let 
\begin{equation}
 {\widehat C}^{(p+1)}  \equiv
 C_{\mu_1...\mu_{p+1}}(Y)\  \partial_{0} 
Y^{\mu_1} ... \partial_{{p}}
Y^{\mu_{p+1}} 
\end{equation}
be the pull-back of the $(p+1)$-form on this worldvolume. The natural
(`electric') coupling is  given by the integral 
\begin{equation}
I_{\rm WZ} =  \rho_{(p)} \int d^{p+1}\zeta\  {\widehat C }^{(p+1)} \ ,
\end{equation}
with $\rho_{(p)}$ the charge-density of the brane. 
Familiar examples are the coupling of 
a  point-particle (`0-brane')  to   
a vector potential, and of a  string (`1-brane')  to   a
two-index antisymmetric tensor. Since the dual of a $(p+1)$-form
potential in ten dimensions is a $(7-p)$-form potential, there exists
also a natural (`magnetic') coupling to a $(6-p)$-brane. The sources
for the field equation  and Bianchi identity  of a $(p+1)$-form are
thus  $p$-branes and $(6-p)$-branes.

Now within type-II  perturbation
theory there are no such elementary RR sources. Indeed,
if a closed-string state  were a source for a  RR $(p+1)$-form, then
the  trilinear coupling
$$
<{\rm closed}\vert\; C^{(p+1)}\; \vert {\rm closed}> \ 
$$
 would not vansih. This is  impossible because 
the coupling  involves  an odd number of left-moving 
(and of right-moving)  fermion emission
vertices, so that the corresponding correlator
vanishes  automatically  on any closed Riemann surface.
What this arguments shows, in particular,  is that
fundamental closed strings do not couple `electrically' to  the Ramond-Ramond
two-form. It is significant, as we will see, that 
in the presence of worldsheet boundaries this simple  argument will
fail.

\begin{figure}
%
%
\begin{center}
\leavevmode
\epsfxsize=14cm
\epsfbox{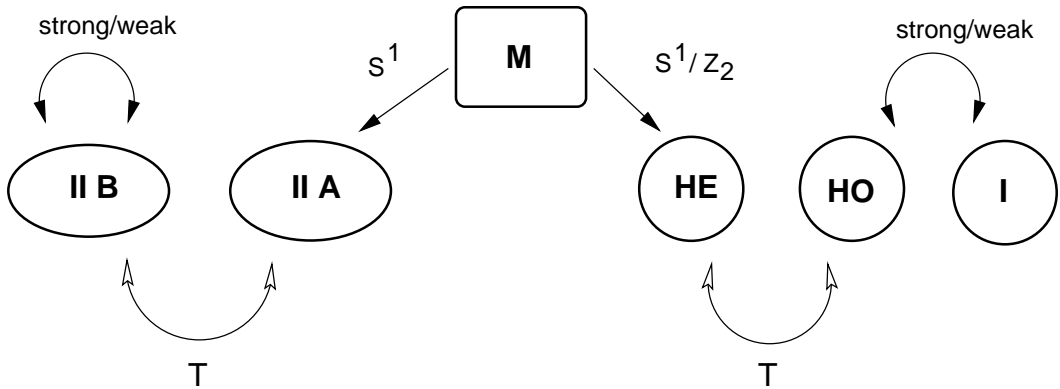}
\end{center}
\caption{ The web of dualities relating the ten-dimensional
  superstring theories and  ${\cal M}$ theory, as described in the text.
\label{fig:web}}
\end{figure}

Most non-pertubative dualities require, on the other hand, the
existence of such elementary RR charges.  The web of string
dualities in nine or higher dimensions,
  discussed in more detail in this volume by Sen
(see also  the other reviews listed  in the introduction), 
has been drawn  in figure 1. The web  holds together the five ten-dimensional
superstring theories, and the eleven-dimensional ${\cal M}$ theory, whose
low-energy limit is
eleven-dimensional supergravity (Cremmer {\em et al} 1978), and which has
a (fundamental~?)  supermembrane (Bergshoeff {\em et al} 1987). 
The black one-way arrows denote  compactifications of
${\cal M}$ theory on the  circle $S^1$,  and  on the  interval $S^1/Z_2$.
In the small-radius limit these are respectively described by
type-IIA string theory (Townsend 1995, Witten 1995), 
and  by the $E_8\times E_8$ heterotic model (Ho\v rava and
 Witten 1996a, 1996b). 
The two-way black arrows identify the strong-coupling limit of one
theory with the weak-coupling limit of another. The type-I and
heterotic SO(32) theories are related in this manner (Witten 1995,
Polchinski and Witten 1996), while the type-IIB theory is self-dual
(Hull and Townsend 1995).
Finally, the two-way white arrows stand for perturbative
T-dualities, after  compactification on an extra  circle (for a review of
T-duality see Giveon {\em et al}  1994).

Consider first the type-IIA theory, whose massless fields are given by
dimensional reduction from eleven dimensions. The bosonic components
of the eleven-dimensional multiplet are the graviton and
 a antisymmetric three-form, and they 
decompose in ten dimensions as follows~:
 \begin{equation}
G_{MN} \rightarrow G_{\mu\nu}, C_\mu, \Phi ; \ \ \
A_{MNR} \rightarrow C_{\mu\nu\rho}, B_{\mu\nu} \ , 
\end{equation}
where $ M,N,R= 0,...,10$.  The eleven-dimensional supergravity has,
however, also Kaluza-Klein excitations  which couple to the 
off-diagonal metric components $C_\mu$. 
Since this is a RR field in type-IIA  theory, duality requires  the existence
of non-perturbative 0-brane charges. 
In what concerns  type-IIB string  theory, its  conjectured
self-duality exchanges the  two-forms
$(B_{\mu\nu}$ and $C_{\mu\nu})$. Since 
fundamental strings are  sources for the Neveu-Schwarz
 $B_{\mu\nu}$, this duality requires  the existence of
non-perturbative  1-branes coupling to  the Ramond-Ramond
$C_{\mu\nu}$ (Schwarz 1995).

  Higher $p$-branes fit  similarly  in the  conjectured web of
dualities. This can be seen more easily after compactification
to  lower dimensions, where
dualities typically mix the various  fields coming
from the Ramond-Ramond and Neveu-Schwarz sectors. For example, 
type-IIA  theory compactified to six dimensions
on a K3  surface is conjectured  to be dual to the heterotic
string compactified on a  four-torus (Duff  and Minasian 1995, Hull and
Townsend 1995, Duff 1995, Witten 1995). 
The latter has extended gauge symmetry at special  points of the
Narain moduli space. On the type-IIA side the maximal abelian gauge
symmetry has gauge fields that descend from the Ramond-Ramond
three-index tensor. These can be enhanced to a non-abelian group only
if there exist charged 2-branes wrapping around shrinking 2-cycles  of
the K3 surface (Bershadsky {\em et al}  1996b). 
A similar phenomenon occurs for Calabi-Yau compactifications
of type-IIB  theory  to four dimensions. 
The  low-energy Lagrangian of Ramond-Ramond fields
has a logarithmic singularity at special (conifold) points
in the Calabi-Yau moduli space. This can be understood as due to
nearly-massless 3-branes, wrapping around shrinking 3-cycles of the
compact manifold, and which have been  effectively integrated out
(Strominger 1995). Strominger's observation was important for two
reasons~: (i) it provided  the first example of a brane that becomes  massless
and can  eventually condense (Ferrara {\em et al}  1995, Kachru and Vafa
1995), and (ii)  in this context the
existence of RR-charged 
branes is not only a prediction of conjectured dualities
-- they {\it have to}  exist  because without them string theory
 would be singular and hence inconsistent.


\section{D-brane tension and charge}

   The only fundamental quanta of string perturbation theory are
elementary strings, so all other  $p$-branes must 
arise as (non-perturbative) solitons. The effective low-energy
supergravities exhibit, indeed, corresponding 
classical solutions (for reviews see 
Duff, Khuri and Lu  1995,   Stelle 1997 and 1998, Youm 1997),
 but these are
often singular and require the introduction of a source. One way to
handle the corrections at the string scale  is to look for
(super)conformally-invariant $\sigma$-models,  a lesson sunk-in from 
the study of string compactifications. Callan {\em et al} (1991a, 1991b) found
such solitonic five-branes in both the type-II and the heterotic
theories. Their branes  involved   only Neveu-Schwarz backgrounds -- 
 being  (`magnetic') sources, in particular,  for the 
two-index tensor  $B_{\mu\nu}$. 
Branes with Ramond-Ramond
backgrounds looked, however,  hopelessly intractable~:  the corresponding
$\sigma$-model would have to involve the  vertex (\ref{eq:vertex}),
which is made out of ghosts and spin fields  and cannot, furthermore, be
written in terms of two-dimensional superfields. 
Amazingly enough,
 these  Ramond-Ramond charged $p$-branes turn  out to admit a much
simpler, exact and universal description 
as allowed   endpoints for open strings, or D(irichlet)-branes  
(Polchinski 1995).

\subsection{Open-string endpoints as defects}
\label{subsec:definition}

The bosonic part of the Polyakov  action for a  free fundamental
  string   in
flat space-time and in the conformal
gauge   reads \footnote{I use the label $a,b\cdots$ both for space-time
  spinors and for the (Euclidean) 
 worldsheet coordinates of a fundamental string --
  the context should, hopefully, help to avoid confusion.}
\begin{equation}
I_{{\rm F}} =  \int_{\Sigma} {d^2\xi \over 4\pi\alpha^\prime}
 \ \partial_a X^\mu
\partial^a X_\mu\ , 
\end{equation}
with $\Sigma$ some generic surface with boundary. For its variation
\begin{equation}
\delta I_{{\rm F}} =  -\int_{\Sigma}\; {d^2\xi \over 2\pi\alpha^\prime}
 \ \delta X^\mu\;
\partial_a\partial^a X_\mu
+ \int_{\partial \Sigma}\; {d\xi^a \over 2\pi\alpha^\prime}
 \ \delta X^\mu
\varepsilon_{a b} \partial^b X_\mu
\end{equation}
to vanish,  the $X^\mu$ must be harmonic functions on the worldsheet,
 and either of the following two conditions must hold on the boundary
$\partial \Sigma$, 
\begin{equation}
\eqalign{ \partial_\perp X^\mu =& 0  \ \ \ {(\rm Neumann)}, \cr 
{\rm  or} \ \ \  
 \delta X^\mu = &0 \ \ \ {(\rm Dirichlet)}\ . \cr}  
\end{equation}
Neumann  conditions  respect   Poincar{\'e} invariance  and
are hence momentum-conserving.
Dirichlet conditions,  on the other hand, 
describe  space-time defects.  They have been  studied in the past
in various guises, for instance  as
sources for partonic behaviour in string theory
 (Green 1991b  and references therein),
as heavy-quark endpoints  (L\" uscher {\em et al}
1980, Alvarez 1981),
and as  backgrounds for  open-string 
compactification (Pradisi and Sagnotti 1989, Ho\v rava 1989b, Dai {\em et
  al} 1989).
 Their  status of  intrinsic non-perturbative
excitations  was not, however, fully appreciated  in these earlier  studies.

A static  defect extending over  $p$ flat spatial dimensions is described
 by the boundary conditions
\begin{equation}
 \partial_\perp X^{\alpha\;=\; 0, \cdots ,\;p} =
 X^{m\;=\; p+1,\cdots ,\;9} = 0 \ 
, \label{eq:plane}
\end{equation}
which  force open strings to move on a $(p+1)$-dimensional (worldvolume)
hyperplane spanning the dimensions $\alpha=0,\cdots, p$.
  Since open strings
do not propagate in the bulk in type-II
theory, their presence is intimately tied to the existence of the
defect, which we will refer to as a D$p$-brane.
Consider  complex radial-time coordinates for the open string -- these  map a
strip worldsheet onto the upper-half plane,
\begin{equation}
z =  e^{\xi^0 + i \xi^1} \ \ \ \ (0<\xi^0<\infty , \ 0<\xi^1<\pi) \ .
\end{equation}
The boundary conditions for the bosonic target-space 
coordinates then take the form 
\begin{equation}
\left. \partial  X^{\alpha} = \overline\partial  X^{\alpha}\ 
 \right|_{\; {\rm  Im}z =0}\ 
 \ \ \ {\rm
  and}\ \ \  
\left. \partial  X^{m} = - \overline\partial  X^{m}\ 
 \right|_{\; {\rm  Im}z =0}\  .
\end{equation}
Worldsheet supersymmetry imposes, on the other hand, the following
boundary conditions on  the worldsheet supercurrents (Green {\em et al}
1987)~:   $J_F =
\epsilon\; {\overline  J_F}$, where 
$\epsilon = +1$ in the Ramond sector, and 
$\epsilon = {\rm sign}({\rm Im}z)$  in the Neveu-Schwarz sector.  
As a result  the fermionic coordinates must obey
\begin{equation}
\left. 
\psi^{\alpha} = \epsilon\; {\bar \psi}^{\alpha}\ \right|_{\; {\rm  Im}z =0}\ 
  \ \ \ {\rm
  and}\ \ \left.  
\psi^{m} = -\epsilon\; \bar \psi^{m}\  \right|_{\; {\rm  Im}z =0}\  . 
\label{eq:fermions}
\end{equation}

To determine the boundary conditions on spin fields, notice that
their  operator-product expansions with the fermions read
 (Friedan {\em et al} 1986)
\begin{equation}
\psi^\mu(z) S(w) \sim (z-w)^{-1/2}\ \Gamma^\mu S(w)
\ , 
\end{equation}
with a similar expression  for  right movers. Consistency with
(\ref{eq:fermions}) 
imposes therefore the  conditions,
\begin{equation}
 \left.  S =  \Pi_{(p)}\;
 \overline S \ \  \right|_{\; {\rm  Im}z =0}\ \ ,
\label{eq:spinbound} 
\end{equation}
where
\begin{equation}
\Pi_{(p)} =   (i\Gamma_{11}\Gamma^{p+1} )(i\Gamma_{11}\Gamma^{p+2} )
 \cdots (i\Gamma_{11}\Gamma^{9} )
\label{eq:oppi}
\end{equation}
is a real  operator which  commutes
 with all   $\Gamma^\alpha$ and
anticommutes  with all  $\Gamma^m$. Since $\Pi_{(p)}$ flips
the spinor chirality for  $p$  even,  only even-dimensional 
D$p$-branes are allowed in type-IIA  theory. For the same reason
type-IIB and type-I theories  allow  only for odd-dimensional D$p$-branes. 
In the type-I  case we  furthermore demand   that 
(\ref{eq:spinbound}) be  symmetric under the interchange 
$S\leftrightarrow  \overline S$.  This implies 
$\Pi_{(p)}^{\ 2} = 1$, which is true only for $p=1,5$ and $9$.
All these facts  are summarized in table 2.


\begin{table}[htp]
\begin{center}
\begin{tabular}{|c|c| }
\hline
 &   \\
type-IIA  & $p$ = 0, 2, 4, 6, 8  \\
 &  \\
 \hline
 &   \\
type-IIB & $p$ = --1, 1, 3, 5, 7, (9)
 \\
 &  \\ 
 \hline
 &  \\
type-I & $p$ = 1, 5, 9
\\
&  \\
  \hline
\end{tabular}
\end{center}
\caption{
The D$p$-branes of the various string theories  
  are (with the exception of the D9-brane)  in one-to-one
correspondence with the `electric'  Ramond-Ramond
potentials of table 1,  and their `magnetic' duals.
 The two heterotic theories have no Ramond-Ramond fields and no
D$p$-branes. }
\end{table}


The case $p=9$ is  degenerate, since it implies that open
strings can propagate in the bulk of space-time. This is only
consistent in type-I theory, i.e. when there are 32 D9-branes and
an orientifold projection. The other 
D$p$-branes listed in the table are in 
one-to-one correspondence with  the `electric'
Ramond-Ramond  potentials of table 1, 
and their `magnetic'  duals. We will indeed  verify that they
couple to these potentials as elementary sources.
The effective action of a D$p$-brane,  with tension
 $T_{(p)}$  and charge density   under the corresponding Ramond-Ramond 
 $(p+1)$-form
$\rho_{(p)}$, reads 
\begin{equation}
 I_{{\rm D}p} \; = \; 
  \int d^{p+1}\zeta \left(  T_{(p)}\; e^{-\Phi}
\sqrt{-{\rm  det}\;{\widehat G}_{\alpha\beta} }
+ \rho_{(p)}\;  
{\widehat C}^{(p+1)} \right)  ,
\label{eq:worldvolume}
\end{equation}
where
\begin{equation}
{\hat G}_{\alpha\beta}= G^{\mu\nu}\partial_\alpha Y_\mu
\partial_\beta Y_\nu
\end{equation}
is the induced worldvolume metric.
The cases  $p=-1,7,8$ are somewhat special. 
The D(-1)-brane sits at a particular  space-time point
and  must be interpreted  as a (Euclidean) instanton with  action
\begin{equation}
 I_{{\rm D}(-1)} \; = \;  
 \left.  T_{(-1)}\; e^{-\Phi}
+ i \rho_{(-1)}\; C^{(0)}\  \right|_{\rm position}  \ .
\label{eq:instanton}
\end{equation}
Its `magnetic' dual, in a sense to be made explicit
below, is the D$7$-brane.  Finally the D$8$-brane is a 
domain wall coupling to the non-propagating nine-form, i.e. 
separating regions with different values of  $H^{(0)}$ (Polchinski and
Witten 1996, Bergshoeff {\em et al} 1996).

The values of $T_{(p)}$
and $\rho_{(p)}$ could  be extracted in principle from
one-point functions on the disk. Following Polchinski (1995) 
we will prefer
to extract them from the interaction energy between two 
static  identical D-branes. 
This approach  will spare us  the technicalities of 
normalizing vertex operators correctly, and will furthermore 
extend  naturally to the study of dynamical 
D-brane interactions (Bachas 1996).


\subsection{Static force:  field-theory calculation}
\label{subsec:force2}

Viewed as solitons of ten-dimensional  supergravity,
 two D-branes interact
by exchanging gravitons, dilatons and antisymmetric tensors.
This is a good approximation, provided their  separation $r$
is large compared to the fundamental string scale.
The  supergravity Lagrangian for the exchanged bosonic fields reads 
 (see Green {\em et al} 1987)
\begin{equation}
I_{\rm IIA,B} = - {1\over 2\kappa_{(10)}^2}
 \int d^{10}x \sqrt{-G}\; \Biggl[
 e^{-2\Phi}
\Bigl( R- 4 (d\Phi)^2  +{1\over 12} (dB)^2
\Bigr)
+\sum {1 \over 2 n!}
 H^{(n)\; 2}  \Biggr]
\label{eq:field}
\end{equation}
where  $n=0,2,4$ for type-IIA theory, $n=1,3$ for type-IIB, while
for the self-dual field-strength $H^{(5)}$ there is no covariant action we
may write down. Since this is a tree-Lagrangian of closed-string
modes, it 
is  multiplied by the usual factor
$e^{-2\Phi}$  corresponding to spherical worldsheet topology. The
D-brane  Lagrangian (\ref{eq:worldvolume}), on the other hand,
is  multiplied by a factor
 $e^{-\Phi}$, corresponding to the topology of the
 disk. The disk is indeed the lowest-genus diagram with 
 a  worldsheet boundary which can  feel the
presence of the  D-brane.
These dilaton
pre-factors have been  absorbed in the terms involving Ramond-Ramond fields
through a rescaling
\begin{equation}
   C^{(p+1)} \rightarrow e^{\Phi} C^{(p+1)} \ .
\end{equation}
A carefull analysis  shows indeed that it is the field strengths
 of the rescaled
potentials which satisfy  the usual Bianchi identity and
  Maxwell equation  when
the dilaton varies  (Callan {\em et al} 
1988,  Li 1996b, Polyakov 1996).

To decouple the propagators of the graviton and dilaton, 
we pass  to  the Einstein metric
\begin{equation}
g_{\mu\nu} = e^{-\Phi/2}  G_{\mu\nu} \ ,
\end{equation}
in terms of which the effective  actions take the form
\begin{eqnarray}
I_{\rm IIA,B}  &=& -{1\over 2\kappa_{(10)}^2}
 \int d^{10}x \sqrt{-g}\ \Bigl[
 R + {1\over 2} (d\Phi)^2 
 +{1\over 12} e^{-\Phi} (dB)^2 \nonumber \\
&&\ \ \ \  + \sum  {1\over 2(p+2)!}
e^{(3-p)\Phi/2} (d C^{(p+1)})^2 \ \Bigr]
\end{eqnarray}
and
\begin{equation}
 I_{{\rm D}p} \;= \;  \int d^{p+1}\zeta\; \left(
 T_{(p)}\;  e^{(p-3)\Phi/4}\;
\sqrt{-{\rm  det}\;{\widehat g}_{\alpha\beta}}\; +\;
  \rho_{(p)}\; \widehat  C^{(p+1)} \right) . 
\end{equation}
To leading order in the gravitational coupling 
the interaction energy comes from  the exchange of
 a single graviton, dilaton or Ramond-Ramond field, and  reads
\begin{equation}
{\cal E}(r)\;{\delta {\rm T}} =
- 2\kappa_{(10)}^2 \int d^{10}x \int d^{10}{\tilde x}
\ \Bigl[  j_{\Phi} \Delta {\tilde j_{\Phi}}
-  j_{C} \Delta  {\tilde j_{C}} +
T_{\mu\nu} \Delta^{\mu\nu,\rho\tau}{\tilde T}_{\rho\tau}
\Bigr]
\end{equation}
Here $j_{\Phi}$, $j_C$ and $T_{\mu\nu}$ are the sources for
the dilaton, Ramond-Ramond form and graviton obtained by linearizing
the worldvolume action for one of the branes, while the
tilde quantities refer to the other brane.
  $\Delta$ 
and $\Delta^{\mu\nu,\rho\tau}$
are the scalar 
and  the graviton propagators in ten 
dimensions, evaluated at
the  argument ($x-{\tilde x}$), and ${\delta {\rm T}}$ the
total interaction time.
To simplify notation, and since only one component of
$C^{(p+1)}$ couples to a  static planar D$p$-brane, we have
 dropped the obvious tensor structure of the antisymmetric field.

The sources for  a static planar defect take the  form
\begin{eqnarray}
j_{\Phi} &=& {p-3\over 4}\; T_{(p)}\; \delta(x^{\perp}) \nonumber \\
j_{C}& =& \rho_{(p)}\;  \delta(x^{\perp}) 
\label{eq:sources}
\end{eqnarray}
$$
T_{\mu\nu} =  {1\over 2}  T_{(p)} \;  \delta(x^{\perp}) \times
\cases{ &$\eta_{\mu\nu}$
\ {\rm if}\ \ $\mu,\nu \leq p$ \cr
& $0$ \ \ \ \ \  {\rm otherwise} \cr}
$$
where
the  $\delta$-function localizes  the defect in transverse space.
The tilde sources are taken identical, except that they are localized
at distance $r$ away in the transverse plane. 
The graviton propagator 
 in the De Donder gauge and in $d$ dimensions reads
(Veltman 1975)
\begin{equation}
 \Delta_{(d)}^{\mu\nu,\rho\tau} = \left(\eta^{\mu\rho}\eta^{\nu\tau}
+ \eta^{\mu\tau}\eta^{\nu\rho} -\frac{2}{d-2}
\eta^{\mu\nu}\eta^{\rho\tau} \right) \;\Delta_{(d)} \ ,
\end{equation}
where
\begin{equation}
\Delta_{(d)}(x) = \int {d^{d} p\over (2\pi)^{d}}
{e^{ipx} \over p^2} \ .
\end{equation}
Putting all this together and  doing some straightforward algebra
we obtain
\begin{equation}
{\cal E}(r) = 2 V_{(p)} \kappa_{(10)}^2
\  [ \rho_{(p)}^2 -  T_{(p)}^2 ]\ 
\Delta_{(9-p)}^E(r)\ ,
\label{eq:sta}
\end{equation} 
where $V_{(p)}$ is the (regularized)  p-brane volume and 
$\Delta_{(9-p)}^E(r)$ is the (Euclidean) scalar
 propagator in $(9-p)$ transverse dimensions. The net force
is as should be expected the difference between Ramond-Ramond
repulsion and gravitational plus dilaton attraction.

\begin{figure}
%
%
\begin{center}
\leavevmode
\epsfxsize=8cm
\epsfbox{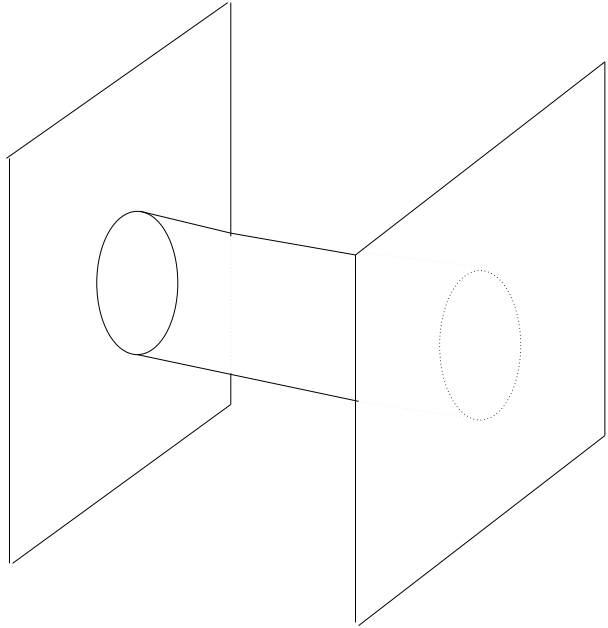}
\end{center}
\caption{  Two D-branes interacting through  the exchange of
a closed string. The diagram has a dual interpretation as
Casimir force due to vacuum fluctuations of open strings.
\label{fig:radish}}
\end{figure}


\subsection{Static force: string calculation}
\label{subsec:force1}

The exchange of all closed-string modes, including the massless
  graviton, dilaton and  $(p+1)$-form,
is given  by the  cylinder diagram of figure 2. 
Viewed   as an annulus, this same diagram also admits a dual
and, from the field-theory point of view,  surprising
interpretation: 
the  two D-branes interact by modifying the
vacuum fluctuations of (stretched) open strings, in the same  way that
two superconducting
plates attract by  modifying  the vacuum fluctuations
of the photon field. It is this simple-minded
duality which
may,  as we will see below,  revolutionize our thinking about space-time.

The one-loop vacuum energy  of oriented  open strings reads
\begin{eqnarray}
& &{\cal E}(r)  =
-{V_{(p)}\over 2} 
 \int {d^{p+1}k\over (2\pi)^{p+1}}\ \int_0^\infty
{dt\over t}\  {\rm Str}\; e^{-\pi t(k^2+ M^2)/2}=
 \nonumber \\
& & \label{eq:static}\\
&=&  - 2\times {V_{(p)} \over 2}\int_0^\infty
{dt\over t}\; (2\pi^2 t)^{-(p+1)/2}\; e^{-r^2 t/2\pi}\ Z(t)
 \ , \nonumber
\end{eqnarray}
where 
\begin{equation}
Z(t) = -{1\over 2} \sum_{s=2,3,4} (-)^s\; {\theta^{4}_s\left(0
  \left.\right|
{it\over 2}\right) \over 
 \eta^{12}\left({it\over 2}\right)}
\end{equation}
is the usual spin structure sum obtained by supertracing over
open-string oscillator states (see Green {\em et al} 1987),
 and we have set $\alpha^\prime =
1/2$. Strings stretching between the two D-branes
  have at the $N$th oscillator level a mass
 $M^2= (r/\pi)^2 + 2N$, so that their  vacuum fluctuations are 
modified when we separate the D-branes. The vacuum energy
of open strings
 with both endpoints on the same  defect is, on the other hand,
$r$-independent and has been omitted. 
Notice also the (important)
 factor of $2$ in front of the second line: it 
accounts for  the two possible orientations of the
 stretched  string,

 The first remark concerning  the above expression,
is that it vanishes by the well-known  $\theta$-function  identity. 
Comparing with eq. (\ref{eq:sta}) we conclude that
\begin{equation}
T_{(p)} = \rho_{(p)} \ ,
\label{eq:BPS}
\end{equation}
so that Ramond-Ramond repulsion cancels exactly the gravitational and
dilaton attraction. As  will be discussed in detail later on, 
 this cancellation of the static force is a consequence of space-time
 supersymmetry. It is similar  to the 
 cancellation of Coulomb repulsion and Higgs-scalar
attraction between  't Hooft-Polyakov monopoles in 
N=4 supersymmetric Yang-Mills (see for example Harvey 1996) .

  To extract the actual value of $T_{(p)}$ we must  
separate in the diagram the exchange of RR and NS-NS closed-string
states. These are characterized by worldsheet fermions
which are periodic, respectively antiperiodic around the
cylinder, so that they  correspond to the $s=4$, respectively
$s=2,3$ open-string spin structures. In the large-separation
limit ($r\to\infty$) we may furthermore expand the integrand
near 
$t\sim  0$~:
\begin{equation}
Z(t) \simeq (8-8)\times \left(\frac{t}{2}\right)^4 + o(e^{-1/t}) \ ,
\end{equation}
where we have here used the standard $\theta$-function asymptotics. 
Using also the
 integral representation
\begin{equation}
\Delta^E_{(d)}(r) = {\pi\over 2} \int_0^\infty dl\;
(2\pi^2 l)^{-d/2}\; e^{-r^2/2\pi l} \ ,
\end{equation}
 and restoring correct mass units we obtain 
\begin{equation}
{\cal E}(r) = V_{(p)}\; (1-1)\; 2\pi (4\pi^2\alpha^\prime)^{3-p}\;
\Delta_{(9-p)}^E(r)\  +\  o(e^{-r/\sqrt{\alpha^\prime}}) \ .
\end{equation}
Comparing with the field-theory calculation we can finally extract
the tension and charge-density of type-II   D$p$-branes ,

\begin{equation}
T_{(p)}^{\; 2} = \rho_{(p)}^{\; 2} = \frac{\pi}{\kappa_{(10)}^2}
(4\pi^2\alpha^\prime)^{3-p}\ .
\label{eq:values}
\end{equation}
These are determined  unambiguously, as should be expected 
for  intrinsic excitations of a  fundamental theory. Notice that in
the type-I theory the above interaction energy should be multiplied by
one half,  because  the stretched open strings are unoriented. The 
tensions and charge densities of type-I D-branes 
are, therefore,  smaller than those of their type-IIB counterparts
by a factor of  ${1\over\sqrt{2}}$.


\section{Consistency and duality checks}
\label{sec:tests}

String dualities and non-perturbative  consistency impose a number of
relations among  the tensions and charge densities of D-branes, which
we will now discuss. We will verify, in particular,  that the
values (\ref{eq:values}) are  consistent with T-duality,
with  Dirac charge quantization, 
as well as  with the existence of an eleventh dimension.
From the string-theoretic point of view, the T-duality
 relations are  the least surprising, since the  symmetry is
automatically built into the genus  expansion. Verifying these relations
is simply a   check of the annulus calculation of the
previous section. That the results obey also the Dirac conditions is
more rewarding, since these test the 
 non-perturbative consistency of the theory.  What
is, however, most astonishing is the fact
 that the annulus calculation `knows' about the
existence of the eleventh dimension.

\subsection{Charge quantization}
\label{subsec:quantization}

Dirac's quantization condition for electric and magnetic charge
(Dirac 1931)
has an  analog for extended objects in higher
dimensions (Nepomechie 1985, Teitelboim 1986a,b).
\footnote{ Schwinger (1968) and Zwanziger (1968)
extended Dirac's  argument to dyons. The
  generalization of their  argument to  higher dimensions 
  involves a subtle sign discussed recently by  Deser {\em et al}
 (1997, 1998).}
Consider a D$p$-brane sitting at the origin, and 
integrate  the field equation of the Ramond-Ramond form 
 over  the transverse space.  Using Stokes'  theorem   one finds
\begin{equation}
\int_{S_{(8-p)}} \ ^*H^{(p+2)}\  =\  2 \kappa_{(10)}^2 \; \rho_{(p)}
\end{equation}
where ${S_{(8-p)}}$ is a (hyper)sphere, surrounding  the defect, in 
transverse space.  
This equation is the analog of  Gauss' law.
Now Poincar{\'e} duality tells us  that
\begin{equation}
\ ^*H^{(p+2)} =  H^{(8-p)}  \simeq  d C^{(7-p)} \ ,
\end{equation}
where  the   potential $C^{(7-p)}$ is not globally defined
because  the D$p$-brane is a source in the
Bianchi identity for $H^{(8-p)}$.
Following Dirac we may define a smooth potential everywhere, 
 except along  a singular (hyper)string which drills a hole in
 $S_{(8-p)}$. The
 hole is  topologically equivalent to the interior of a
hypersphere $S_{(7-p)}$. 
  These facts are easier to
visualize in three-dimensional space, where a point defect creates a
string singularity which drills a disk out of a  two-sphere, 
 while a string defect creates a sheet singularity which drills a
 segment out of a  circle,  as  in figure 3.

\begin{figure}
%
%
\begin{center}
\leavevmode
\epsfxsize=10cm
\epsfbox{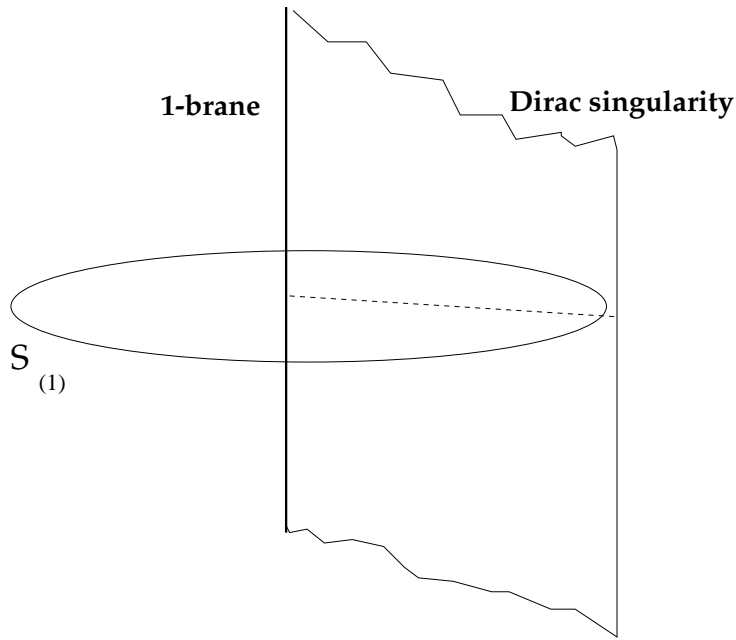}
\end{center}
\caption{  A 1-brane  creates
a 3-index ``electric'' field $H^{(3)}$. Electric flux
in d=4 space-time dimensions  is given
by an integral of the dual vector over a circle $S_{(1)}$.
The `magnetic' potential is  a scalar field,  coupling to point-like 
(Euclidean) instantons, and jumping discontinuously  across the depicted 
 sheet singularity.
\label{fig:ra}}
\end{figure}


 The Dirac  singularity  is dangerous because  a Bohm-Aharonov 
experiment involving $(6-p)$-branes might detect it. 
Indeed, the wave-function of
a  $(6-p)$-brane transported 
around the  singularity  picks a phase
\begin{equation}
{\rm Phase} =  \rho_{(6-p)} \int_{S_{(7-p)}} C^{(7-p)} \ = \ 
 \rho_{(6-p)} \int_{S_{(8-p)}} H^{(8-p)}\ . 
\end{equation}
For the (hyper)string to be  unphysical, this phase must be an
integer multiple of $2\pi$.
 Putting together equations (4.1-4.3) we thus find the condition
\begin{equation}
{\rm Phase}=   2\kappa_{(10)}^2
\rho_{(p)}\rho_{(6-p)} = 
 2\pi n \ .
\label{eq:Dirac}
\end{equation} 
The  charge densities (\ref{eq:values})  satisfy this
condition with $n=1$. 
D-branes are  therefore the  minimal  Ramond-Ramond  charges
allowed in the theory,
 so one may conjecture  that there are no others.

Dirac's  argument is strictly-speaking valid only 
for $0\leq p\leq 6$ \footnote{The D3-brane is actually also special,
  since it couples to a self-dual four-form.}. In order to extend it 
 to the pair $p=-1,7$,   note  that a D7-brane creates a 
 (hyper)-sheet singularity across which  the Ramond-Ramond scalar,
 $C^{(0)}$, jumps  discontinuously  by an amount
$2\kappa_{(10)}^2 \rho_{(7)}$.  Dirac quantization ensures that the
exponential of the  (Euclidean) instanton action 
(\ref{eq:instanton}) has no discontinuity across the sheet, whose 
presence cannot therefore  be detected by non-perturbative physics.
It is the four-dimensional analog of this special case that is, as a
matter of fact, 
illustrated in figure 3.

A final comment 
concerns the type-I theory, where  the extra factor of ${1\over\sqrt{2}}$
in the charge densities seems to violate the quantization condition.
 The puzzle is resolved by
the observation (Witten 1996b)
 that the   dynamical five-brane excitation consists of  a {\it pair}
of coincident  D5-branes, so that 
\begin{equation}
\rho_{(1)}^{\;I}
 =  \sqrt{\frac{\pi}{2 \kappa_{(10)}^2} } \; (4\pi^2\alpha^\prime) 
\ \ \ {\rm and} \ \ \ 
\rho_{(5)}^{\;I} = 2\times 
\sqrt{\frac{\pi}{2\kappa_{(10)}^2}} \; 
(4\pi^2\alpha^\prime)^{-1}\ .
\end{equation}
This is consistent with heterotic/type-I duality, as well as  with the fact
that the orientifold projection removes the collective coordinates of
a single,  isolated D5-brane (Gimon and Polchinski 1996).


\subsection{T-duality}
\label{subsec:T}

T-duality is a discrete  gauge  symmetry  of string  theory,
 that transforms  both the background fields and the 
 perturbative (string)
 excitations around them (see Giveon {\em et al} 1994).
The simplest context in which it occurs 
is compactification of type-II theory on a circle.
The general expression for the  compact (ninth)  coordinate of a
closed  string is 
\begin{eqnarray}
 z\partial X^9  =   {i\over 2} \left( \frac{n_9\alpha^\prime}{R_9} +
 m_9R_9\right)
 +i\sqrt{\alpha^\prime  \over 2}
 \sum_{k\not= 0} {a^9_k}\;  z^{-k}\nonumber  \\
 {\bar  z}{\overline \partial} X^9  = 
 {i\over 2} \left( \frac{n_9\alpha^\prime}{ R_9} -  m_9R_9\right)
+ i\sqrt{\alpha^\prime  \over 2}
 \sum_{k\not= 0} {{\widetilde   a}^9_k}\; {\bar  z}^{\;-k}
\label{eq:modes}
\end{eqnarray}
Here $n_9$ and $m_9$ are the quantum numbers corresponding to 
momentum and winding,
and  $z= e^{\xi^0+i\xi^1}$ with $0\leq \xi^1 < 2\pi$.
A  T-duality transformation
inverts  the radius of the circle,  interchanges  winding with
momentum numbers, and flips the sign of right-moving oscillators~:
\begin{equation}
R_9^\prime = \frac{\alpha^\prime}{R_9}\ , \ \ \  
 ( n_9^\prime, m_9^\prime )= (m_9, n_9)\ \ {\rm and}\ \  
{\widetilde  a}^{9\ \prime}_k= - {\widetilde   a}^{9}_k .
\label{eq:T}
\end{equation}
It also shifts the expectation value of the dilaton, so as to leave
the nine-dimensional Planck scale unchanged,
\begin{equation}
{R_9^\prime \over \kappa_{(10)}^{\prime\ 2}} =
{R_9\over \kappa_{(10)}^{\ 2}}\ .
\label{eq:Eins}
\end{equation}

The transformation (\ref{eq:T}) can be thought of  as a 
(hybrid)  parity operation   
restricted  to  the antiholomorphic worldsheet sector~:  
\begin{equation}
{\overline \partial} { X}^{9\  \prime} =  -{\overline \partial}{X}^9 \ .
\end{equation}
Since the  parity operator in spinor space is  $i\Gamma^9\Gamma_{11}$, 
bispinor fields will transform accordingly as follows:
\begin{equation}
H^\prime =  i H\; \Gamma^9\Gamma_{11}  \ .
\end{equation}
Using the gamma-matrix identities of section 2, we may rewrite
this relation in  component form, 
\begin{equation}
H^\prime_{\mu_1 ... \mu_n} =   H_{9\; \mu_1 ... \mu_n}  \ \ \ {\rm
  and}\ \ \ 
H^\prime_{9\;\mu_1 ... \mu_n} = -H_{\mu_1 ... \mu_n}\ , 
\label{eq:TRR}
\end{equation}
for any  $\mu_i \not =9$.  T-duality exchanges 
therefore even-$n$  with
odd-$n$ antisymmetric field strengths,
 and hence also type-IIA with type-IIB
backgrounds. Consistency requires that it also transform
even-$p$ to odd-$p$ D-branes and vice versa.

To see how this comes about let us   consider a D$(p+1)$-brane
wrapping  around the ninth dimension. We concentrate on the ninth
coordinate of an  open string living on this D-brane. It can be
expressed as the sum of the holomorphic and anti-holomorphic pieces
(\ref{eq:modes}), with   an extra factor  two  multiplying
the zero modes becuase  the open string is
parametrized by  $\xi^1\in [0,\pi ]$. The Neumann boundary condition
$\partial X^9 = \overline\partial X^9$ at real $z$, forces furthermore
the identifications
\begin{equation}
a_k = {\widetilde a_k}\ , \ \ \ {\rm and} \ \ \ m_9 = 0\ .
\end{equation}
 This is consistent with the fact that
open strings can   move freely along the ninth dimension on the
D-brane,  but cannot wind.

Now  a T-duality transformation flips the sign of the antiholomorphic
piece, changing  the  Neumann
to a Dirichlet condition, \footnote{For general curved backgrounds with
  abelian isometries this has  been discussed by Alvarez {\em et al}
  1996,   and by Dorn and Otto 1996.} 
\begin{equation}
a_k^\prime = - {\widetilde a_k}^\prime\ , \ \ \ {\rm and} \ \
 \ n_9^\prime = 0\ .
\end{equation}
The wrapped D$(p+1)$-brane is thus
transformed, in the dual theory, 
 to a D$p$-brane localized in the ninth dimension
 (Ho\v rava 1989b, Dai {\em et al} 1989, Green 1991a).
 Open strings cannot move along this  dimension anymore,
but since their endpoints are fixed on the defect they can now wind. 
The inverse transformation is also true: a 
D$p$-brane,  originally transverse to the ninth  dimension,  transforms
to  a wrapped  D$(p+1)$-brane in the dual theory.
All this is  compatible with the
transformation (\ref{eq:TRR}) of  Ramond-Ramond fields, to  which the various
D-branes couple. Furthermore, since a gauge transformation
 should not change the (nine-dimensional)
tension  of the defect, we must have
\begin{equation}
2\pi R_9\; T_{(p+1)} = T_{(p)}^\prime \ .
\end{equation}
Using the formulae  (\ref {eq:T}-\ref {eq:Eins}) one can 
check  that the D-brane tensions indeed verify this T-duality constraint. 
Conversely,  T-duality plus  the minimal Dirac quantization condition 
fix unambiguously the expression (\ref{eq:values}) for the D-brane tensions.

\subsection{Evidence for d=11}
\label{subsec:tests}

  The third  and most striking set of relations are
 those derived from the conjectured duality between  
 type-IIA string theory and  ${\cal M}$  theory compactified on a circle
(Witten 1995, Townsend 1995).
The eleven-dimensional supergravity couples  consistently to a
supermembrane (Bergshoeff {\em et al}  1987),  and has furthermore  a 
(`magnetic') five-brane (G\" uven 1992) with a non-singular
 geometry (Gibbons {\em et al}
1995). After compactification on the circle there exist also
Kaluza-Klein modes, as well as  a Kaluza-Klein monopole given  by
the   Taub-NUT$\times {\rm R}^7$ space  (Sorkin 1983, Gross and Perry 1983).  
The correspondence between these excitations and the various branes on
the type-IIA side is shown in table 3.
The missing entry in this table is the 
eleven-dimensional counterpart of  the D8-brane, which 
  has not yet been identified
 (for  a recent attempt see
Bergshoeff {\em et al} 1997).
 The problem is that massive type-IIA
supergravity (Romans 1986), which prevails on one side of the wall
(Polchinski and Witten 1996, Bergshoeff {\em et al} 1996),
 seems to have no ancestor in eleven  dimensions 
(Bautier {\em et al} 1997, Howe {\em et al} 1998).

\begin{table}[htp]
\begin{center}
\begin{tabular}{|c|c||c|c|}
 \hline
& & &  \\
{\bf tension}&  {\bf type-IIA}  & {\bf  ${\cal M}$  on $S^1$}
  &  {\bf tension } \\
&  & &  \\
\hline \hline
 & & & \\
$\displaystyle{\sqrt{\pi}\over  \kappa_{(10)}} 
\textstyle (2\pi\sqrt{\alpha^\prime})^3 $
 & D0-brane &  K-K excitation & $\displaystyle {1\over  R_{11}}$  \\
 & &  &\\
 \hline
 & &  & \\
$T_F= (2\pi\alpha^\prime)^{-1}$ & string  & wrapped membrane  &
$2\pi R_{11} \displaystyle \left( {2 \pi^2 \over \kappa_{(11)}^{\
      2}}\right)^{1/3} $ 
  \\
 & & & \\ 
 \hline
 & & & \\
$\displaystyle{\sqrt{\pi}\over  \kappa_{(10)}}\textstyle
 (2\pi\sqrt{\alpha^\prime})$ & D2-brane  & membrane  &
$ T_2^M= \displaystyle \left( {2 \pi^2 \over \kappa_{(11)}^{\
      2}}\right)^{1/3} $  \\
& &  &\\
  \hline
 & & & \\
$\displaystyle{\sqrt{\pi}\over  \kappa_{(10)}}\textstyle 
 (2\pi\sqrt{\alpha^\prime})^{-1}$
 & D4-brane   & wrapped five-brane &
$R_{11} \displaystyle \left( {2 \pi^2 \over \kappa_{(11)}^{\
      2}}\right)^{2/3} $
 \\ 
& & & \\ 
 \hline
 & & & \\
$\displaystyle {\pi \over \kappa_{(10)}^{\ 2}}\textstyle
 (2\pi\alpha^\prime)$ & NS-five-brane & five-brane  & 
$ \displaystyle {1\over 2\pi} \left( {2 \pi^2 \over \kappa_{(11)}^{\
      2}}\right)^{2/3} $
 \\
& &  &\\
  \hline
 & & & \\
$\displaystyle{\sqrt{\pi}\over  \kappa_{(10)}}\textstyle
 (2\pi\sqrt{\alpha^\prime})^{-3}$
 & D6-brane   & K-K monopole & 
$ \displaystyle  {2 \pi^2 R_{11}^{\ 2}  \over \kappa_{(11)}^{\  2}} $ \\ 
& & & \\ 
 \hline
 & & & \\
 $\displaystyle{\sqrt{\pi}\over  \kappa_{(10)}}\textstyle 
 (2\pi\sqrt{\alpha^\prime})^{-5}$
 & D8-brane   &~? &~?  \\ 
& & & \\ 
 \hline
\end{tabular}
\end{center}
\vskip 0.4cm
\caption{
Correspondence of BPS excitations of type-IIA string theory,  and of
$\cal M$
theory compactified on a  circle. Equating tensions and the
ten-dimensional Planck scale on both sides gives seven relations for
two unknown parameters. Supersymmetry and
consistency imply three Dirac quantization conditions, leaving us with
two independent checks of the conjectured duality. }
\end{table}

  Setting aside the D8-brane, let us consider the tensions of the
remaining excitations listed in table 3. 
The tensions  are  expressed in terms of
$\kappa_{(10)}$ and the Regge slope on the type-IIA side,
and in terms of $\kappa_{(11)}$ and the compactification radius
on the ${\cal M}$-theory side.  To
compare sides  we must identify the ten-dimensional
Planck scales, 
\begin{equation}
{1  \over  \kappa_{(10)}^{\ 2} } = {2\pi R_{11}\over \kappa_{(11)}^{\ 2}}
 \ .
\end{equation}
Equating  the fundamental string tension ($T_F$)
 with the tension  of a wrapped membrane fixes also
 $\alpha^\prime$ in terms of
eleven-dimensional parameters. This leaves us with  five  consistency
checks of the conjectured duality, which are indeed explicitly verified.

 How much of this truly  tests the eleven-dimensional origin of
string  theory? To answer the question we must first understand how the
entries on the ${\cal M}$-theory side of table 3 are  obtained.
Because of the scale invariance of the  supergravity
equations, the  tensions of the classical membrane and 
fivebrane solutions are a priori arbitrary.  
Assuming minimal Dirac quantization, and the BPS equality of mass and
charge,  fixes the  product 
\begin{equation}
2\kappa _{(11)}^{\; 2} T_2^M T_5^M = 2\pi \ .
\end{equation}
An argument fixing each of the tensions separately was 
first given by Duff, Liu and Minasian (1995) and further developped by 
de Alwis (1996,1997) and   Witten (1997a).  \footnote{ See also 
 Lu (1997), Brax and Mourad (1997, 1998) and  Conrad (1997).}
It uses the   Chern-Simons term  of  the eleven-dimensional Lagrangian,
\begin{equation}
I_{11d} = -{1\over 2\kappa_{(11)}^{\ 2}}\; \int d^{11}x\ 
\left[  \sqrt{-G}\; \left( R + {1\over 48} (dA)^2 \right)
 + {1\over 6 } A\wedge
  dA\wedge dA  \right] \ ,
\end{equation}
where  $A \equiv {1\over 3~!} A_{MNR}\; dx^M\wedge dx^N\wedge dx^R$
 is the three-index antisymmetric form encountered already
in section 2.
In a nutshell, the  coefficient of this 
Cherm-Simons  term is fixed by supersymmetry (Cremmer {\em et al} 1978),
but in the presence of electric and magnetic sources it is 
 also subject to an independent quantization condition. 
\footnote{The quantization of
 the abelian  Chern-Simons term in the presence of a magnetic
source  was first discussed in 2+1 dimensions (Henneaux and Teitelboim
1986, Polychronakos 1987). } 

 Let me describe the argument in the simpler context of 
 five-dimensional Maxwell theory with a (abelian)  Chern-Simons term, 
\begin{equation}
I_{5d}^{\rm MCS}
 = -{1\over 2\kappa_{(5)}^2} \int d^5x \; \left({1\over 4} F^2 +
  {k \over 6}\; A\wedge
  F\wedge F \right)\ .
\label{eq:maxwell}
\end{equation}
Assume  that the theory has both elementary  electric charges $q$
(coupling through  $I_{WZ} = q \int A_\mu dx^\mu $), and  dual  
minimally-charged magnetic strings. If we compactify the 
fourth spatial dimension on a circle of radius $L$, 
the effective four-dimensional action  reads
\begin{equation}
I_{4d}^{\rm MCS}
 = -{1 \over 2 \kappa_{(4)}^2} \int d^4x \; \left({1\over 4} F^2 +
  {1\over 2} (da)^2 + 
  {k \over 2}\; a\;
  F\wedge F \right)\ , 
\label{eq:maxwell1}
\end{equation} 
where $\kappa_{(5)}^2 = 2\pi L \kappa_{(4)}^2$  and  $a = A_4$. 
The scalar field $a$  must be
 periodically identified, since
under a large gauge transformation 
\begin{equation}
a \to  a + {1\over qL}\ . 
\label{eq:shift}
\end{equation}
Such a shift changes, however, the $\theta$-term of the
four-dimensional Lagrangian, and is potentially observable through the
Witten effect, namely  as a  shift in  the electric charge of a  magnetic
monopole (Witten 1979). This latter is a  magnetic string  wrapping
around the compact fourth dimension.  To avoid an immediate
 contradiction we must require  that
the  induced charge be an integer multiple of $q$, so that it can  be
screened by  elementary charges bound to the monopole.

 In order to  quantify this requirement,  consider the
 $\theta$-term resulting from the shift (\ref{eq:shift}). 
In the background of a monopole field it  will give  rise to an
interaction (Coleman 1981)
\begin{equation}
-{ k\over 2 \kappa_{(4)}^2 qL}
 \int d^4x \  F_{r0}^{\ *}F_{r0}^{(\rm monopole)} =
{ 2\pi^2  k  \over \kappa_{(5)}^2 q^2} \int dt\; A_0   \  ,
\label{eq:induced}
\end{equation}
where we have here integrated by parts and used the monopole equation
  $\partial_r^{\  *}F_{r0}^{(\rm monopole)} 
= (2\pi/q)\; \delta ^{(3)}(\vec r)$.
The  interaction (\ref{eq:induced}) 
 describes  precisely the Witten effect, i.e. the
fact that the magnetic monopole has acquired
a non-vanishing  electric charge. 
Demanding that the induced charge 
 be an integer multiple of  $q$   leads, finally, to 
  the quantization condition
\begin{equation}
 q^3 = {2\pi^2 k  \over  \kappa_{(5)}^2 n }\ . 
\end{equation}
This is the sought-for relation between
  the coefficient of the (abelian) Chern-Simons term and  the
elementary electric charge of the theory.

  Let us apply now the same reasonning to  
  ${\cal M}$-theory.
Compactifying to eight dimensions on a three-torus gives an 
effective eight-dimensional theory with both electric and magnetic
membranes. The latter are the wrapped five-branes of  ${\cal
  M}$-theory, which may   acquire
an  electric charge through a generalized Witten effect. 
 Demanding that a large
gauge transformation induce  a charge that can be screened by
elementary membranes 
leads to the quantization condition
\begin{equation}
(T_2^M)^3 = {2\pi^2    \over  \kappa_{(11)}^2 n }\ .
\end{equation}
This relates the electric  charge density or
membrane tension, $T_2^M$, to the
coefficient, $k=1$, of the Chern-Simons term.
The membrane tension predicted by duality  corresponds  to the
maximal allowed case  $n=1$.

  We can finally return  to our  original question~: How much evidence
for the existence of an eleventh dimension in string theory
 does  the `gedanken data' of table 3  contain?
Note first that  Dirac
quantization relates the six tensions pairwise.  Furthermore, since the
maximal non-chiral 10d supergravity is unique, it must contain a
$B\wedge H^{(4)} \wedge H^{(4)}$ term obtained from the 11d 
Chern-Simons term  by dimensional reduction. An argument similar to
the one described above can then be used to
fix the product $T_{(2)}^{\ 2}T_F$  of D2-brane and fundamental-string
tensions. Thus, supersymmetry
and consistency determine (modulo integer ambiguities) all but two of
the tensions of  table 3, without any reference either to the ultraviolet
definition of the theory or to the existence of an
eleventh dimension. We are therefore left 
 with a single truly independent check
 of the conjectured duality, which we can take to be  the relation
\begin{equation}
 T_{(0)} T_F = 2\pi\; T_{(2)} \ .
\label{eq:single}
\end{equation}    
This is a trivial geometric identity in ${\cal M}$-theory, which had
no a priori reason to be satisfied from the ten-dimensional viewpoint.

  The sceptic reader may find that a single test  constitutes little
evidence for the duality conjecture.
\footnote{
 To be sure, the existence of threshold bound states of D-particles --
the Kaluza-Klein modes of the supergraviton --  constitutes  further, a priori
independent,  evidence for the duality conjecture (Yi 1997, Sethi and Stern
1998, Porrati and Rozenberg 1998).}
The above discussion, however, 
underscores what might be the main  lesson of the `second string
revolution'~:  the ultimate theory may  be  unique  {\it precisely}
because reconciling quantum mechanics and gravity
is such a  constraining enterprise.


\section{D-brane interactions}

  D-branes in supersymmetric configurations exert no net static force  on
  each other, because (unbroken)  supersymmetry ensures that the
  Casimir energy of open strings is zero. Setting the branes in
  relative motion (or rotating them) breaks generically all the 
  supersymmetries, and leads to velocity- or orientation-dependent
  forces. We will now extend Pochinski's calculation  to study such
  D-brane interactions. Some suprising new insights come from
 the close relationship between brane
dynamics and supersymmetric gauge theory -- a theme that will be
recurrent in this and  in the subsequent  sections.
Two results
of particular  importance, because they lie at the heart of the
  M(atrix)-model conjecture of Banks {\em et al} (1997),   are 
the dynamical appearance of the eleven-dimensional Planck length, 
and the simple scaling with distance  of the leading 
low-velocity interaction of D-particles.    
Since space-time supersymmetry plays a key role in our 
discussion, we will first describe in some more detail  the general 
BPS configurations of D-branes.


\subsection{BPS configurations}
\label{subsec:BPS}

 A planar static D-brane is a BPS defect that leaves half of
the  space-time  supersymmetries  unbroken.
This follows  from the equality
$T_{(p)} = \rho_{(p)}$, and   the 
(rigid) supersymmetry algebra, appropriately extended to take into
account $p$-brane charges (de Azcarraga {\em et al} 1989, see  Townsend
1997 for a detailed discussion).  
Alternatively, we can draw this  conclusion  from  a worldsheet
point of view. On a closed-string  worldsheet the thirty-two space-time
 supercharges are given by contour integrals of the fermion-emission
 operators,
\begin{equation}
Q = \oint {dz\over z} \; S \ \ \ {\rm and}\ \ \ 
{\overline Q} = - \oint {d\bar z\over \bar z} \; {\overline 
S} \ \ .
\end{equation}

\begin{figure}
%
%
\begin{center}
\leavevmode
\epsfxsize=10cm
\epsfbox{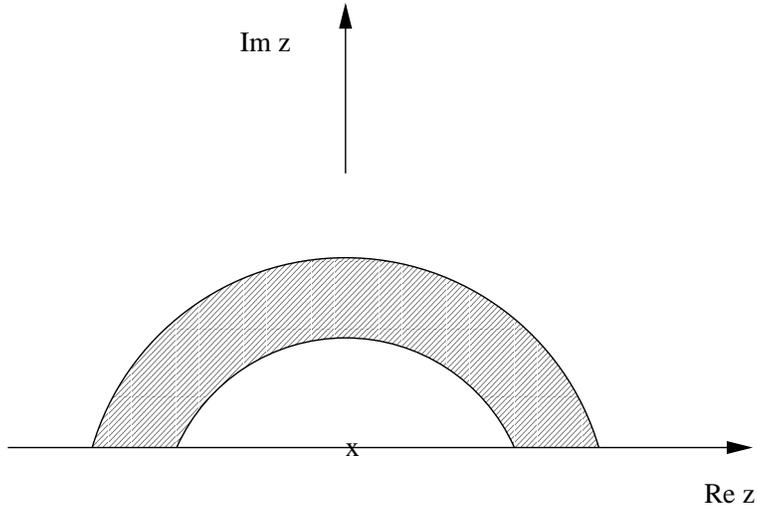}
\end{center}
\caption{  The semi-circles are two snapshots of an open
string at fixed radial time $\xi^0 = {\rm log}|z|$. 
A charge is conserved when  its time variation can be
expressed as a holomorphic plus  antiholomorphic contour integral around the
shaded region in the upper complex plane. This means  that the
contributions of the linear segments on the worldsheet boundary
should  vanish. 
\label{fig:bps}}
\end{figure} 

\noindent  Holomorphicity allows us to deform the integration
contours,   picking  (eventually) extra  contributions only from  points
where  vertex operators have been inserted. This leads to 
supersymmetric Ward identities for the perturbative
 closed-string S-matrix in  flat  ten-dimensional space-time.

  Now in  the background of a D$p$-brane we must also  define the action of
the (unbroken) 
 supercharges on the  open strings. The corresponding integrals, 
 at fixed radial time $\xi^0$,  run over a semi-circle as  in figure 4.
Moving  the  integration  to a   later time, 
is allowed only if the contributions of  the worldsheet boundary 
vanish. This is the case for the sixteen linear combinations
\begin{equation}
Q +  \Pi_{(p)}{\overline Q} =
 \int_0^\pi  d\xi^1 \; ( S + \Pi_{(p)} {\overline 
S}) \ \ ,
\label{eq:unbroken}
\end{equation}
for which the  holomorphic and antiholomorphic
pieces add up to zero  on the real axis  by virtue of
the boundary  conditions (\ref{eq:spinbound}). The remaining sixteen
supersymmetries  are broken spontaneously by the D$p$-brane, and cannot
thus be realized linearly within the perturbative string expansion.

 Consider next a background with two planar static D-branes, to which
 are  associated  two operators,  $\Pi_{(p)}$ and 
$\widetilde \Pi_{(\widetilde p)}$. These operators
depend on the  orientation, 
but not on the position, of the branes. More explicitly, we can put
equation (\ref{eq:oppi})  in  covariant form 
\begin{equation}
\Pi_{(p)} = - {i^{p+1} \over (p+1)!}
 \; \omega^{(p)}_{\mu_{0}\;\cdots\; \mu_p}\;
\Gamma^{\mu_{0}\;\cdots\; \mu_p}
\end{equation}
where
\begin{equation}
\omega^{(p)} \equiv {1\over (p+1)!}\; \omega^{(p)}_{\mu_0\;\cdots\;\mu_p}\;
 dY^{\mu_0}\wedge\cdots\wedge dY^{\mu_p}
= \sqrt{ -\widehat g}\; d\zeta^0\wedge\cdots\wedge d\zeta^p
\end{equation}
is the (oriented)  volume  form of the D$p$-brane, 
 and we have
done some simple $\Gamma$-matrix rearrangements.
There is of course a similar expression for the tilde brane. 
In the background of these two  D-branes, the linearly-realized
 supercharges are  a subset of (\ref{eq:unbroken}), namely 
\begin{equation}
{\cal P} (Q + \Pi_{(p)} \overline Q)
= \int_0^\pi  d\xi^1 \; {\cal P} ( S + \Pi_{(p)} {\overline 
S}) \ \ ,
\end{equation}
with ${\cal P}$ an appropriate 
 projection operator.  Demanding that
the corresponding contour integrals cancel out on a worldsheet
boundary that is stuck on the tilde brane leads to the condition
\begin{equation}
 {\cal P}\widetilde \Pi_{(\widetilde p)} = {\cal P} \Pi_{(p)} \ ,
\label{eq:proj}
\end{equation}
which admits a non-vanishing  solution if and only if 
\begin{equation}
{\rm det} \left({\bf 1} - \widetilde
 \Pi_{(\widetilde p)}\Pi_{(p)}^{-1}\right) = 0 \ .
\end{equation}
The number of unbroken supersymmetries is the number of zero
eigenvalues of the above  matrix.  Every extra  D-brane  and/or
orientifold imposes of course one  extra condition, which has to be
satisfied simultaneously.

  A trivial solution to these BPS equations is given by  two 
(or more) identical, parallel D$p$-branes  at arbitrary separation
$r$. This background  preserves  sixteen supersymmetries  and has, of
course, a $r$-independent vacuum energy,
  consistently with the cancellation of
forces found by Polchinski.  Flipping the
orientation of one brane sends $\Pi_{(p)} \to - \Pi_{(p)}$, thus breaking
all space-time supersymmetries. The resulting configuration describes
a brane and an anti-brane, attracting both gravitationnally, and
through  Ramond-Ramond exchange. In the 
force calculation of section
3.3, this amounts to reversing the sign of the $s=4$ spin structure,
i.e. of the GSO projection for the stretched open string. The
surviving Neveu-Schwarz ground state
becomes, in this case,  tachyonic at a critical separation
$r_{\rm cr} = \sqrt{2\pi^2\alpha^\prime}$,  beyond which
 the attractive  force between the brane and the anti-brane  diverges 
(Banks and Susskind 1995, Arvis 1983).

   Other   solutions to the BPS conditions can be found with two
orthogonal D-branes. For such a configuration 
\begin{equation}
\widetilde \Pi_{(\widetilde p)}\Pi_{(p)}^{-1} =
\pm  \prod_{m\in\; p\sqcup  
 {\widetilde p}} \Gamma^m\ ,
\end{equation}
where $p\sqcup {\widetilde p}$ denotes  the set of dimensions
spanned by one or other of the branes  but not  both, and the overall
sign depends on the choice of orientations.
 The eigenvalues of the above
operator depend only on  the even number ($d_{\perp}$)  of dimensions in 
$p\sqcup {\widetilde p}$.\  For $d_{\perp} = 4n+2$ the eigenvalues are all
purely  imaginary, and supersymmetry is completely broken. 
For  $d_{\perp} = 4$ or $8$, on the other hand,
 half of the eigenvalues are $+1$, so
 eight of the supersymmetries are linearly-realized in the  background.
Examples of $d_{\perp} =4$
configurations (for early discussions see 
Bershadsky {\em et al} 1996a, Sen, 1996, Douglas 1995)
 include a D4-brane and a D-particle, a D5-brane
and a parallel D-string, or two  completely transverse  D2-branes.
Examples  of 
$d_{\perp} =8$ configurations  are a D8-brane and a D-particle, or 
two completely transverse  D4-branes. As the reader  can verify easily,
all   configurations with the
same value of $d_\perp$ can be (at least formally) related by the
T-duality transformations of section 4.2.

    It will be useful later on to know  the spectrum of
 an open string stretching  between two orthogonal D-branes.  Such a
 string has $d_\perp$ coordinates obeying mixed (DN) boundary conditions~:
Neumann  at one  endpoint  and Dirichlet  at the other.
 A bosonic DN coordinate  has a half-integer mode expansion,
 while its fermionic partner is
 integer modded in the Neveu-Schwarz sector and half-integer modded in
 the Ramond sector. Using the standard expressions for the subtraction
 constants of integer or half-integer modded fields, we find the
mass formula  (in units $2\alpha^\prime =1$), 
\begin{equation}
 M^2\; = \; \left({r\over \pi}\right)^2 +  {2 N_{\rm osc}}\; + \;
\cases{ {{d_\perp/  4} - 1} & \ NS\cr  
\ \ \ 0 & \  R\cr}\ ,
\end{equation}
with $N_{\rm osc}$ the sum of the  oscillator frequencies. 
Furthermore,  Neveu-Schwarz  and Ramond states
are  spinors of  $SO(d_\perp)$
and  $SO(1,9- d_\perp)$ -- the two  maximal Lorentz subgroups
that such a  brane configuration could  leave unbroken. 
Note in particular that for
 $d_\perp = 4$ the massless states have  the
content of a six-dimensional  hypermultiplet,
 while for $d_\perp =8$ the only
massless state is a two-dimensional 
 (anti)chiral fermion, which is a singlet of
the unbroken chiral (8,0) supersymmetry (Banks,
Seiberg and Silverstein 1997, Rey 1997).

 There exist also  BPS configurations with 
D-branes at arbitrary angles (Berkooz
{\em et al} 1996). 
A solution, for instance,  of equation (\ref{eq:proj}) with 
 eight unbroken supersymmetries  is given  by 
$ {\cal P}  = {1\over 2} ({\bf 1} -
  \Gamma^6\Gamma^7\Gamma^8\Gamma^9)\ $ , and 
\begin{equation}
 \widetilde\Pi_{(\widetilde p)} \Pi_{(p)}^{-1} \; = \; - 
 ({\rm cos}\theta\; \Gamma^6
+ {\rm sin}\theta\; \Gamma^8)({\rm cos}\theta\; \Gamma^7+
 {\rm sin}\theta\; \Gamma^9)\; \Gamma^6  \Gamma^7 \ .
\end{equation}
It  describes 
two identical D-branes, one of which spans the dimensions (67) 
and is transverse to the dimensions (89), 
whereas the second has undergone a (relative) 
unitary rotation in the
  $\CC^2$ plane $(x^6+ix^7, x^8+ix^9)$.
 The case $d_{\perp}=4$ discussed above  corresponds
to the special angle $\theta = {\pi\over 2}$.
A more exotic example with  six unbroken supersymmetries can be obtained
by a  rotation that preserves a quaternionic structure (Gauntlett
{\em et al} 1997). For a review of  BPS  configurations of  
 intersecting and/or overlapping branes 
 see Gauntlett (1997).

\subsection{D-brane scattering}
\label{subsec:scatt}

 The velocity-dependent forces between D-branes  can be analyzed by
 calculating  the 
semi-classical phase shift for two moving  external sources. I will here
 follow the original calculation (Bachas 1996) for two identical
 D$p$-branes in the Neveu-Ramond-Schwarz formulation. The same results
 can be obtained in the light-cone boundary-state formalism
 (Callan and Klebanov 1996, Green and
 Gutperle 1996,  Billo {\em et al} 1997),   and
can be furthermore extended to  different D-branes (Lifschytz 1996),
 non-vanishing  worldvolume  fields (Lifschytz 1997, Lifschytz and
 Mathur 1997, Matusis 1997), orbifold backgrounds (Hussain {\em et al} 1997), 
type-I theory (Danielsson and Ferretti 1997), 
and to study spin-dependent interactions (Morales {\em et al} 1997,1998).

\begin{figure}
%
%
\begin{center}
\leavevmode
\epsfxsize=7cm
\epsfbox{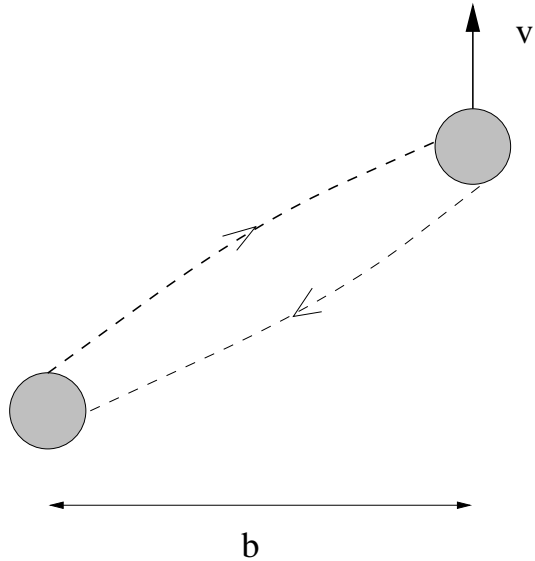}
\end{center}
\caption{ Two D-particles scattering with relative
  velocity $v$ and impact parameter $b$. The broken lines depict  a
  virtual pair of oriented strings 
   being stretched by the relative motion. The imaginary part of
  the phase shift gives the probability that these virtual strings
  materialize.
\label{fig:scatter}}
\end{figure}

   Consider two identical parallel D$p$-branes, one of which is at
rest, while the second is  moving 
with  velocity $v$ and impact parameter $b$, as shown in  figure 5. 
It is convenient to define  the boost parameter corresponding to this 
relative motion,
\begin{equation}
v\equiv  {\rm tanh} (\pi\epsilon)  \ .
\end{equation}
Thinking of the D-brane  interaction as
Casimir force leads us to study the spectrum of
an open string stretched between these  two external  sources. 
If  the motion
 is along the ninth dimension, the coordinates
$X^{1,\cdots,8}$ retain their  conventional mode expansions.
The mode expansion
of the  light-cone coordinates $X^\pm = (X^0\mp  X^9)/\sqrt{2}$, on the other
hand, is  modified to  
\begin{equation}
X^\pm = -i\sqrt{\alpha^\prime\over 2} \sum_{k=-\infty}^{\infty}
\left[ {a_k^\pm \over k\pm i\epsilon}\; z^{-k\mp i\epsilon}
+ { a_k^\mp \over k\mp i\epsilon}\; {\bar z}^{\; -k\pm i\epsilon}  \right] \
.
\end{equation}
It is indeed easy to  verify  that $X^0$ and $X^9$ obey, 
respectively,  Neumann  and Dirichlet conditions at $\xi^1=0$,
so that one  end-point of the open  string is fixed on the static
D$p$-brane. Furthermore, 
 $X^\pm(\xi^0,\pi) =e^{\pm \pi\epsilon} X^\pm(\xi^0,0)$,
so that the other  endpoint  is
boosted with velocity $v$, consistently with the fact that it is fixed
on the moving D$p$-brane. 
The mode expansions of the fermionic light-cone coordinates can be
derived similarly with the result
\begin{equation}
\psi^\pm = \sqrt{\alpha^\prime}\;
\sum_k \psi_k^\pm\; z^{-k\mp  i\epsilon} \ , \ \ 
\bar\psi^\pm = \sqrt{\alpha^\prime}\;
\sum_k  \psi_k^\mp\;  \bar z^{\;-k\pm i\epsilon} \ ,
\end{equation}
where $k\in \ZZ$ in the Ramond sector of the open string, while
 $k\in \ZZ+{1\over 2}$ in the  Neveu-Schwarz sector.

The relevant  feature in  the above expressions  is the shift of all
oscillator frequencies by an amount $\pm i\epsilon$. Similar
expansions arise in the twisted sectors of an orbifold, with
$i\epsilon$ replaced by a (real) rotation angle. 
Using the standard formulae for the partition functions of free
 massless fields   with twisted
boundary conditions  we find (here $2\alpha^\prime =1$)
\begin{equation}
\delta (b,v)  =
  - 2\times {V_{(p)} \over 2 }\int_0^\infty
{dt\over t}\; (2\pi^2 t)^{-p/2}  e^{-b^2 t/2\pi}\ Z(\epsilon, t)
 \ , 
\label{eq:phaseshift}
\end{equation}
where  
\begin{equation}
Z(\epsilon, t)  = -{1\over 2} \sum_{s=2,3,4} (-)^s \;
 {\theta_s\left({\epsilon t\over 2} \left|\right. {it\over 2} \right)
\over \theta_1\left({\epsilon t\over 2} \left|\right.  {it\over 2} \right)}
\;
{\theta^{3}_s\left(0
  \left.\right|
{it\over 2}\right) \over 
 \eta^{9}\left({it\over 2}\right)}
\  .
\label{eq:Z}
\end{equation}
Expressions (\ref{eq:phaseshift}--\ref{eq:Z})
 generalize  Polchinski's calculation to the case of moving D-branes.
As a check of  normalizations notice that in the leading $v\to 0_+$
approximation the above result  reduces 
 correctly to the quasi-static phase shift
\begin{equation}
\delta(b, v) \simeq  \int_{-\infty}^{\infty} d\tau\; {\cal
  E}(\sqrt{b^2+v^2\tau^2})\ + \cdots ,
\end{equation}
with ${\cal E}(r)$ the  static interaction energy,  eq.
(\ref{eq:static}). This follows from the fact that for small (first)
argument the function 
 $\theta_1$ vanishes linearly, with   $\theta_1^\prime(0\vert \tau) =
2\pi\eta^3(\tau)$.  Of course, since the D-branes feel no static force, this
leading quasi-static phase shift is  zero. 

  The supergravity, 
$b\to\infty$,  limit of the phase shift can be obtained from the 
$t\to 0$ corner of the  integration region. It is to this end
convenient to first put the partition function, 
 using  Jacobi's  identity, in the simpler form
(Green and Gutperle  1996), 
\begin{equation}
Z(\epsilon, t)  = 
 {\theta_1^4 \left({\epsilon t\over 4} \left|\right. {it\over 2} \right)
\over \theta_1\left({\epsilon t\over 2} \left|\right.  {it\over 2} \right)
 \eta^{ 9}\left({it\over 2}\right) }
\  .
\label{eq:Jacobi}
\end{equation}
With the help of the modular transformations
\begin{equation}
\theta_1 \left(-{\nu\over\tau} \left|  -{1\over\tau} \right.
 \right) = \sqrt{i\tau}\; 
e^{i\pi\nu^2/\tau}\;  \theta_1(\nu\vert\tau)\ \ , 
\ \ \ \eta(-{1/\tau}) = \sqrt{-i\tau}\; \eta(\tau) \ , 
\end{equation}
as well as of  the product representations (here $q=e^{2 i\pi\tau}$)
\begin{equation}
{\theta_1 (\nu\vert\tau)\over \eta(\tau)}  = 
2 q^{1\over 12}\; {\rm sin}(\pi\nu)\prod_{n=1}^\infty
 (1-q^{n} e^{2\pi i\nu})
(1-q^{n} e^{-2\pi i\nu})\  
\end{equation}
and 
\begin{equation}
 \eta(\tau) = q^{1\over 24}\;
\prod_{n=1}^\infty (1-q^n) , 
\end{equation}
we can easily extract the $t\to 0$ behaviour  of the partition function. 
This leads to the following  asymptotic behaviour for the
phase shift, in the limit $b\to\infty$:
\begin{equation}
\delta   \simeq - {V_{(p)}
\over (2\pi\sqrt{\alpha^\prime})^{p}}\; \Gamma\left(3-{p\over 2}\right)
\left( {4\pi\alpha^\prime\over b^2}\right)^{3-p/2} 
{{\rm sinh}^4(\pi\epsilon/2)\over {\rm sinh}(\pi\epsilon)}
  +\;  \cdots  \ ,
\end{equation}
where the   corrections
come   from the  exchange of massive closed strings, and hence  fall off
exponentially with distance.
It is a simple (but tedious) exercise to 
recover   the above
result by repeating the  calculation  of section 3.2,
with  one of the two  external sources boosted to a moving frame.
 Alternatively, 
this  result can be compared to  the classical action for
geodesic motion in the appropriate supergravity
 background (Balasubramanian and Larsen 1997).


\subsection{The size of  D-particles}
\label{subsec:size}

  That the  string calculation should reproduce  the supergravity result
at sufficiently large impact parameter  is  reassuring, 
but hardly surprising. 
A more interesting question to address is what happens if we try to probe
a D-brane at short, possibly substringy  scales.  
 In order to answer this question let us 
note that the partition function $Z(\epsilon, t)$ has poles along the
integration axis,  at 
$\epsilon t/2 = k\pi$  with $k$ any  odd positive integer.  These
correspond to
 zeroes  of the trigonometric sine  in the product
representation of $\theta_1$. As a result the phase shift acquires  an
imaginary (absorptive) part, equal to  the sum over the 
positions of the poles of $\pi$ times the residue of the integrand.
A straightforward calculation gives (Bachas 1996)
\begin{equation}
{\rm Im}\;\delta\; = \; \sum_{\rm multiplets}
{{\rm dim}(s) \over 2}  \sum_{k\ {\rm odd}}  {\rm exp}
\left[ - {2\pi\alpha^\prime k\over \epsilon}
 \left(   {b^2\over (2\pi\alpha^\prime)^2}
   +  M(s)^2 \right)  \right]\ ,  
\end{equation}
where the sum
runs over all supermultiplets of
 dimension ${\rm dim}(s)$ and oscillator-mass
 $M(s)$ 
 in the  open-string spectrum,  and we have restricted our attention
to D-particles, i.e. we have set $p=0$. This result has a simple
interpretation~: as the two D-particles move away from each other,
they transfer continuously their energy to any open strings that
happen to  stretch  between them (see figure 5).
 A virtual pair of open strings can thus materialize from
the vacuum and stop completely,  or slow down the  motion. 
\footnote{Since the D-branes are extremely heavy at weak string
coupling, the back reaction is a higher-order effect.
For a discussion of D-brane recoil see
Berenstein {\em et al} (1996), and Kogan {\em et al} (1996).}
The phenomenon  is T-dual to the more familiar
pair production in a background
electric field, whose rate  in open string theory has been calculated
earlier  by Bachas and Porrati (1992). It is worth stressing that
this imaginary  part cannot  arise from  the exchange of any  finite number 
of closed-string states.

The onset of this dissipation  puts a lower limit on the distance scales
probed by  the scattering, 
\begin{equation}
b\  \succ\  \sqrt{\epsilon/T_F} \ ,
\end{equation}
where $T_F = (2\pi\alpha^\prime)^{-1}$ is the fundamental string
tension, and the symbol $\succ$ stands for `sufficiently larger
than'. In the ultrarelativistic regime  $v\to 1$, so that  $\epsilon\simeq
-{1\over 2\pi} {\rm log}({1-v^2}) \gg 1$. The D-particle  behaves  in
this limit as a black absorptive disk,  of area  much bigger than string scale
and growing logarithmically with energy.
This  typical Regge  behaviour characterizes also
the high-energy scattering
of fundamental strings (see for example Amati {\em et al} 1987, 1989). 
To probe substringy distances 
we must  consider the opposite regime of  low velocities, 
$\epsilon\simeq v/\pi \ll 1 $. The stringy halo is  not excited, in this
regime, all the way down to impact parameters $b \succ \sqrt{v/\pi T_F}$.
Quantum mechanical uncertainty ($\Delta x\Delta p \succ 1$) puts, on
the other hand,  an independent  lower limit
\begin{equation}
b\;  T_{(0)} v \ \succ \  1 \ .
\end{equation}
Saturating  both
bounds  simultaneously  gives 
\begin{equation}
b_{\rm min}^3  \sim {1\over T_F T_{(0)}}  \sim {1\over T_{(2)}} \ ,
\end{equation}
where we have used here the tension formula (\ref{eq:single}).
 We thus conclude that the
 dynamical size of D-particles is comparable to
 the  (inverse cubic root of the)
 membrane tension,   i.e. to  the 
 eleven-dimensional Planck scale of M-theory!
Since this is smaller than string length at weak string-coupling,
 perturbative string theory does not  
capture all the degrees of freedom at short scales. 
\footnote{ One should
contrast this with  the example of 
 magnetic monopoles in  N=2, d=4 Yang-Mills theory,
whose size is comparable to the Compton wavelength of the fundamental quanta.
Thus, even though  monopoles  are very heavy at weak coupling,
 the high-energy behaviour of the theory
 is still correctly captured by the (super)gauge bosons.}

   The fact that D-branes are much smaller than the  fundamental strings
at weak string  coupling  
 was  conjectured early on 
  by Shenker (1995).  The
appearance of the eleven-dimensional Planck scale in the
matrix quantum mechanics of D-particles
 was first noticed by Kabat and Pouliot (1996) and by Danielsson
and Ferretti (1996).  A systematic analysis of the above
 kinematic regime, and of the validity of the  approximations,
was  carried out  by Douglas {\em et al} (1996).  
Needless to say that this small  dynamical scale of D-particles
cannot be seen by using  fundamental-string probes --
 one cannot  probe a needle with a jelly
pudding, only  with a second  needle!  This is confirmed by explicit
calculations of closed-string scattering off target D-branes
(see Hashimoto and Klebanov 1997, 
 Thorlacius 1998 and references therein).

 One other  striking feature of the low-velocity dynamics of D-particles
follows from an analysis of  the real part of the phase shift. 
Expanding expressions (\ref{eq:phaseshift}, \ref{eq:Jacobi}) for
$\epsilon\simeq v \to 0$, and for any impact parameter, we find 
\begin{equation}
\delta(b,v) \simeq -{\epsilon^3}\;{\left( 2\pi^2 \alpha^\prime
\over b^2\right)^3}  + o(\epsilon^7) \ .
\label{eq:flat}
\end{equation}
Notice that $\delta$  must
flip sign under time reversal \footnote{This would  not be  true 
if the scattering  branes carried  electric and magnetic charge.}
and is hence an odd function of velocity, and that
the  interaction time blows up as $1/\vert v\vert$ because two slow
particles stay longer in the vicinity of  each other.
 The generic form of the
low-velocity expansion therefore is~: 
$\delta(v,b) = \delta_0(b)/v + \delta_1(b)v + \delta_2(b)v^3 + \cdots
$.
 Comparing with  eq. (\ref{eq:flat}) we conclude that, not only the
static,  but also the $o(v^2)$ force between two  D-particles  is zero.
Since the $o(v^2)$
scattering of heavy solitons
can be described by geodesic motion in the  moduli space of
zero modes (Manton  1982), what we 
learn  is that the moduli
 space of D-particles is
(at least  to this order of the genus expansion) completely flat.
Furthermore, as first  recognized clearly by Douglas {\em et al} (1997),
the leading $o(v^4)$ interaction has the same power-law dependence 
on impact parameter
in the supergravity regime ($b\gg \sqrt{\alpha^\prime}$) 
as  at substringy scales ($b\ll \sqrt{\alpha^\prime}$).  
Both of these facts are a result of   space-time supersymmetry. As will
become, indeed,  clear in the following section, our
phase-shift calculation could  be rephrased as
a  one-loop calculation of the effective quantum action 
for a vector  multiplet,  in a theory with 
sixteen  unbroken  supercharges. Velocity is related by supersymmetry to
the field strength, so that  the $o(v^{2k})$  force  between 
D-branes can be read off the $2k$-derivative terms
in  the quantum  action. Using  helicity-supertrace formulae it can
be shown that,  at one-loop order, the two-derivative terms are
not corrected,  while
the only contributions to the four-derivative terms come from 
 short (BPS) supersymmetry mutliplets
(Bachas and Kiritsis 1997). Since all excited states of an open string
are non-BPS, this explains  why the leading $o(v^4)$ interaction has  a
trivial dependence on the string scale.

 This result implies  that the (matrix) quantum mechanics of D-particles,
 obtained by truncating the open-string theory to its lightest modes,
 captures correctly  the leading  $o(v^4)$  supergravity interactions.
It was, furthermore, shown recently that these interactions
 are not modified  by higher-loop and
  non-perturbative corrections 
(Paban {\em et al} 1998, see also 
Becker and Becker 1997, Dine and Seiberg 1997, Dine {\em et al} 1998).
 This is an important ingredient of the
 conjecture by Banks {\em et al} (1997), which will not be pursued
 here any further. We will instead go back now and discuss 
 the classical worldvolume actions of D-branes.


\section{Worldvolume  actions}

Although the  full dynamics of a  soliton  cannot be separated from the
field theory in which it belongs, its  low-energy dynamics
can be approximated by quantum mechanics in the moduli-space
of zero modes. For an extended $p$-brane defect,  the zero modes
give rise to massless worldvolume fields,  and the quantum mechanics
becomes a $(p+1)$-dimensional field theory. 
Similar considerations apply to a D$p$-brane, whose long-wavelength
dynamics we will analyze in this section. Two striking
features of this analysis are (i) the natural emergence of a noncommutative
space-time, and (ii) the power of the combined constraints of
T-duality and Lorentz invariance.

\subsection{Noncommutative geometry}
\label{subsec:Noncom}

The  perturbative excitations of a static planar 
D$p$-brane are described by an  open string theory, interacting with
the  closed strings
in the bulk. The  Dirichlet boundary conditions do not modify the
usual spectrum of the open string, but force its 
center-of-mass momentum to lie along the D-brane. 
The low-energy excitations  make up, therefore, 
a  vector supermultiplet  
dimensionnally-reduced from ten down to  $(p+1)$  dimensions,
\begin{equation}
A^\mu(\zeta^\beta)
 \rightarrow   A^{\alpha\; =\; 0,...,p}(\zeta^\beta)\; ,
\; Y^{m\; =\; p+1,...,9}(\zeta^\beta) \ \ 
. \end{equation}
The worldvolume scalars $Y^m(\zeta^\beta)$ are the transverse space-time
coordinates of the D$p$-brane, i.e. the Goldstone modes of broken
translation invariance. There are no physical degrees of freedom in
the longitudinal coordinates, which in the natural  `static' gauge are
used to parametrize the worldvolume,  $Y^\alpha = \zeta^\alpha$. The extra
physical bosonic excitations of the open string correspond instead to a
worldvolume-vector gauge field -- a feature that characterizes all
D-branes, and which was overlooked  in the earlier supersymmetric
`brane scans' (see Duff  1997).

\begin{figure}
%
%
\begin{center}
\leavevmode
\epsfxsize=10cm
\epsfbox{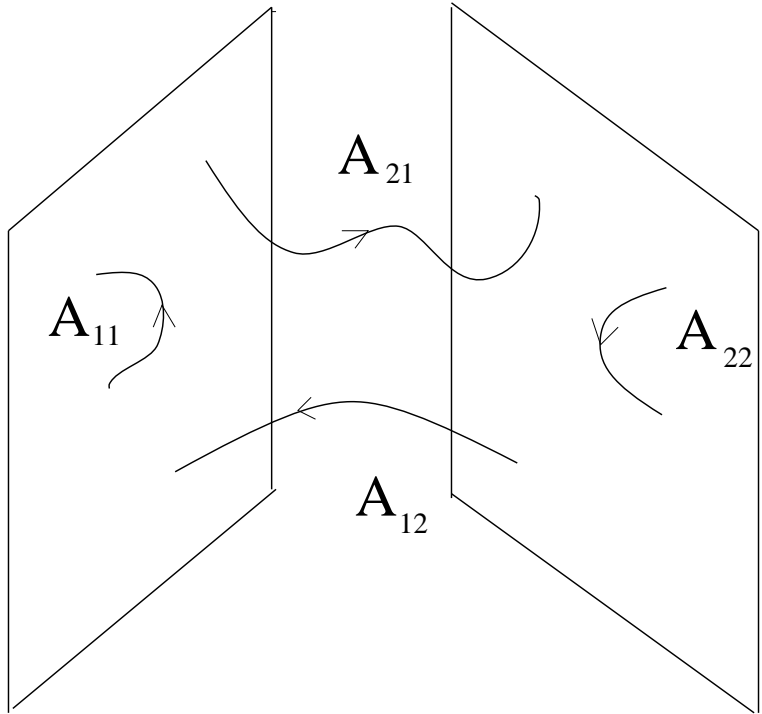}
\end{center}
\caption{
A  D-brane sandwich, and the
 four types of oriented open strings giving rise to massless states
 in the coincidence limit. Each multiplet $A_{ij}$ contains a
 worldvolume vector and $9-p$ worldvolume scalars. These latter are
 the non-commuting transverse coordinates of the D-branes.
\label{fig:nonabelian}}
\end{figure}

 That much can  be in fact deduced from an analysis of the low-energy
supergravity solutions.  String theory
becomes, however, important  when one considers multiple, closely-spaced 
  D-branes.
In addition to the massless vector multiplets describing the
dynamics  of each defect,
there are now extra potentially-light fields
corresponding to the (ground states of)  open strings stretching  between  
different D-branes.  
In the simplest case
of $n$  parallel identical   D$p$-branes, 
the ensuing low-energy field theory
is  a dimensionnally-reduced 
maximally-supersymmetric Yang-Mills theory as above, but with a
non-abelian gauge group
\begin{equation}
U(n) \simeq U(1)_{\rm CM} \times SU(n)_{\rm relative} \ .
\end{equation}
This is, indeed, the low-energy limit of an oriented open-string
theory with a Chan-Paton index $i=1,\cdots n$, 
labelling the $n$  possible string  endpoints. 

The special case $n=2$  is illustrated in figure 6. 
The scalar fields of the $U(1)$  vector multiplet, 
 $Y_{\rm CM}^m = (Y_{11}^m+Y_{22}^m)/2 $, 
are the  transverse center-of-mass coordinates  of the `sandwich',
while the relative motion is described by 
(matrix-valued)  coordinates  in the Lie algebra of $SU(2)$.
The scalar potential, $V\sim {\rm tr}\; [Y^m, Y^r]^{\;2}$, is flat for any
mutually commuting expectation values  $<Y^m>$. These correspond
precisely to the arbitrary positions of the two D-branes, consistently
with the fact that the net static force vanishes. 
 At non-zero separation, 
the complex vector multiplet $A_{12}+ i A_{21}$ acquires a
mass, and  the $SU(2)$  gauge theory is in a  spontaneously-broken,
Coulomb  phase.  It is intriguing that
the  permutation  of the brane positions  is an element of the Weyl
subgroup of $SU(2)$ -- quantum indistinguishability of the excitations
is thus part of the local, gauge symmetry in this picture.
The   non-abelian  nature  of the D-brane
coordinates,  first recognized  clearly  by Witten (1996a),
puts in a precise context
earlier more general  ideas about  the possible role of
noncommutative geometry in physics
(see  for instance Connes 1994, Madore 1995).

  In the type-I theory, the above  $U(n)$  vector multiplet
 is truncated by the
orientifold projection.  The  projection must  antisymmetrize
the Chan-Paton indices of the worldvolume vector and symmetrize those
of the worldvolume scalars, or vice versa. The reason is that the
corresponding vertex operators, 
 \begin{equation}
V^\alpha = 
\int d\xi^a\;
  \partial_a  X^\alpha\; e^{ip \cdot X}\;   , \ \ \ 
V^m  =  {1\over 2\pi\alpha^\prime} \int d\xi_a\;
  \epsilon^{ab}  \partial_b  X^m \; e^{ip\cdot X}  \; ,
\label{eq:vertices}
\end{equation}
have opposite parity under worldsheet orientation reversal.
Tadpole cancellation forces an  antisymmetric projection for the
9-branes, giving the standard $SO(32)$  gauge group.
Consistency of the operator-product expansions then requires
(Gimon and Polchinski 1996) a $SO(n)$  gauge group for the D-strings  and a 
$USp(n)$  group for the D5-branes, with the  worldvolume scalars
 in the symmetric, respectively 
antisymmetric,  $n\otimes n$  representations. A single D-string,
in particular, has no worldvolume gauge fields, consistently with the
fact that it is dual  to the heterotic string
(Polchinski and Witten 1996).   Likewise
a single D5-brane has no transverse coordinates -- the minimal dynamical
excitation, dual to the heterotic five-brane,  is a 
 pair of D5-branes with a worldvolume $USp(2)\simeq SU(2)$ gauge field
(Witten 1996b).

Figure 6  summarizes in  itself many  of the  new insights brought
by D-branes. The  light states of stretched open strings, generically
invisible in the  effective supergravity,
 are the important degrees of freedom in various
settings.  They are responsible, in particular,
 for the thermodynamic properties of
near-extremal black holes (Strominger and Vafa 1996,
Callan and Maldacena 1996), and for the richness of the D-particle
spectrum which lies at the heart  of  the (M)atrix-model conjecture 
(Banks {\em et al} 1997). Furthermore, the realization of  
supersymmetric gauge field theories as worldvolume theories has
led to an improved  understanding of the former through 
brane constructions ( Hanany and Witten 1997,  Banks {\em et al} 1996,
Elitzur {\em et al}  1997a,1997b), while more  recently  the connection with
supergravity has raised  new hopes of solving certain large-$n$
superconformal gauge theories in the planar,  't Hooft
limit (Maldacena 1997,  Gubser {\em et al} 1998, Witten 1998).

\subsection{Dirac-Born-Infeld  and Wess-Zumino terms}
\label{subsec:action}

The effective  action of a  D-brane,  used in the force
calculation of
section 3,  can be generalized
 to take into acount the dynamics of  the worldvolume gauge field, 
and the coupling to arbitrary supergravity backgrounds. 
The action for a single D-brane can be written  as
\begin{equation}
 I_{{\rm D}p} =  \int d^{p+1}\zeta \ \left(  {\cal
 L}_{\rm DBI} +  
 {\cal L}_{\rm WZ} + \cdots  \right)  \ ,
\label{eq:Daction}
\end{equation}
where the Dirac-Born-Infeld
 and Wess-Zumino (or Chern-Simons) lagrangians
are given by 
\begin{equation}
{\cal L}_{\rm DBI}\;  = \;  T_{(p)}\; e^{-\Phi}\;
 \sqrt{ - {\rm det} \left( {\widehat G}_{\alpha\beta}
 + {\widehat
    B}_{\alpha\beta} + 2\pi\alpha^\prime F_{\alpha\beta} \right) }\ , 
\label{eq:BI}
\end{equation}
and
\begin{equation}
{\cal L}_{\rm WZ}  =  \;  T_{(p)}  \;\left. 
 {\widehat C }\wedge  e^{ 2\pi\alpha^\prime F} 
\wedge {\cal G}
\ \ \right|_{(p+1)-{\rm form}} \ .
\label{eq:WZ}
\end{equation}
Here    
$ {\widehat B}_{\alpha\beta}$ is the pull-back of the Neveu-Schwarz
 two-form, 
\begin{equation}
{\widehat C} \equiv  \sum_n\  {1\over n!}\;
{\widehat C}_{\alpha_1 \cdots  \alpha_n}\; d\zeta^{\alpha_1}
\wedge \cdots\wedge d\zeta^{\alpha_n} \ 
\end{equation}
is the sum over all electric and magnetic  RR-form potentials, pulled
back to the worldvolume of the D-brane, and 
$F=dA$ is the
worldvolume field-strength two form,
 normalized so that the  coupling on a boundary
 of the fundamental-string worldsheet is  $\oint A_\alpha dX^\alpha$.
The geometric  part of the Wess-Zumino
 action reads
\begin{equation}
{\cal G} \; =  \;
\sqrt{{\cal A}( {\cal T})/{\cal A}( {\cal N}) } \; =
\; 1 - {(4\pi^2 \alpha^\prime)^2 \over 48}\; \Bigl[ \; p_1( {\cal T}) -  
 p_1( {\cal N})\; \Bigr]     + \cdots
\label{eq:G}  
\end{equation}
where ${\cal T}$ and ${\cal N}$ are  the tangent and normal
bundles of  the brane, $ {\cal A}$  is the
appropriately-normalized
`roof genus',\footnote{The conventional normalization amounts to
choosing  units  $ 4\pi^2 \alpha^\prime =1$,  in which 
all type-II D-branes have the same tension $\sqrt{\pi}/\kappa_{(10)}$.
The roof genus is also frequently denoted ${\widehat{\cal A}}$, but I
here reserve the use of hats to denote pullbacks on  the worldvolume.
}
and
$p_1$ is the  first Pontryagin class 
 (see for instance Milnor and
Stasheff 1974,  Eguchi {\em et al}  1980, or Nakahara 1990
 for definitions). The next term in the expansion of (\ref{eq:G}) is
 an eight-form, whose presence in the D-brane action has not been
 explicitly checked.
All multiplications in ${\cal L}_{WZ}$,
 including those in the Taylor expansions of the
square root and of the exponential,
must be understood in the sense of forms -- 
 what one integrates  in
the end is the coefficient of the $d\zeta^0\wedge\cdots\wedge
d\zeta^p$ term in the expansion.  Strictly-speaking, 
since the RR potentials cannot be 
globally defined in the presence of D-branes, one must 
 use the fact
that $ ( e^{ 2\pi\alpha^\prime F} 
\wedge {\cal G} -1) $ is an exact form, and integrate by parts 
 to  express all but Polchinski's 
coupling  in terms of  the RR field-strengths.
Note, finally,  that since ${\cal T}\oplus {\cal N} = 
{\widehat {\cal S}}$,  the pullback of the space-time tangent
bundle, we can use the multiplicative property of the roof genus, 
\begin{equation}
{ {\cal A}}({\cal T})\wedge  { {\cal A}}({\cal N})\; = \;
{ {\cal A}}({\widehat{ \cal S}})\; =\; 1 +
{(4\pi^2\alpha^\prime)^2\over 192\pi^2}\; {\rm tr}\; ({\widehat {\cal R}}
\wedge {\widehat {\cal R}}) + \cdots 
\ , 
\end{equation}
to  trade the dependence on either ${\cal T}$ or ${\cal N}$ 
for dependence on  the (pulled back)  target-space
curvature two-form,   $\widehat {\cal R}_{\alpha\beta}\;
d\zeta^{\alpha} \wedge d\zeta^{\beta}$.

The Dirac-Born-Infeld lagrangian 
is a generalization  of the geometric volume of the brane trajectory,
 in the presence of
Neveu-Schwarz antisymmetric tensor and worldvolume gauge fields
(Leigh 1989). It was first derived in the context of type-I string theory in
ten dimensions (Fradkin and
Tseytlin 1985). The Wess-Zumino
lagrangian, on the other hand, generalizes Polchinski's 
coupling of  D$p$-branes   to  Ramond-Ramond $(p+1)$-forms.  The
gauge-field dependence was derived  by Li (1996a) and Douglas (1996),
and   the gravitational terms  for trivial normal bundle by
 Bershadsky {\em et al} (1996) and  Green {\em et
al} (1997). The extension to non-trivial normal bundles was given in
special cases  by Witten (1997b) and Mourad (1997), and
more generally by  Cheung and Yin (1997) and  Minasian and Moore
(1997).   Unlike ${\cal L}_{\rm WZ}$, which is related as we will see to
anomalies  and is thus believed  to be exact, the 
`kinetic' action is   known to 
receive corrections involving
acceleration terms and derivatives of the field-strength background
(Andreev and Tseytlin 1988, Kitazawa 1987). These  reflect the
non-local nature of the underlying open-string theory.
The fermionic completion of the action 
(\ref{eq:BI}-\ref{eq:WZ}), compatible with
space-time supersymmetry and with worldvolume
 $\kappa$-symmetry,  has been derived by several authors
(Bergshoeff and Townsend 1997,
Cederwall {\em et al}  1997a.1997b, Cederwall 1997,
Aganagic {\em et al } 1997a,1997b, 
Abou Zeid and Hull 1997), but will not be discussed in this lecture.

  The generalization of this action to multiple D-branes is
non-trivial.  The  transverse fluctuations
$Y^m$,  the `tangent frame'  $\partial_\alpha Y^\mu$  used to pullback 
tensors to the worldvolume, and the  field strength $F_{\alpha\beta}$, all
take now their values  in the Lie algebra of $U(n)$.   The
tree-level action,  obtained  from the disk diagram, must 
be given by a single trace
in the fundamental representation of the gauge group, but
the ordering of the various terms is a priori ambiguous.
Things  simplify   considerably if  the supergravity backgrounds
do not depend on  the  coordinates $x^m$ that are
transverse to the unperturbed D-branes.  
\footnote{ The more general case
has been considered by Douglas (1997) and Douglas, Kato and Ooguri
(1997).}  T-duality reduces in
this case the problem to that of finding the 
non-abelian extension  of the 
gauge-field action only. This is straightforward 
for  ${\cal L}_{\rm WZ}$, in which  we need only  make the replacement
\begin{equation}
 e ^{2\pi\alpha^\prime F} \to 
{\rm tr}_n\;   e ^{2\pi\alpha^\prime F} \ .
\end{equation}
The proper non-abelian generalization  of the Born-Infeld action,
on the other hand, is not known. The leading quadratic term in the
$\alpha^\prime$-expansion of this action is unambiguous,   thanks to
the cyclic property
of the trace. The ordering ambiguities in the next-to-leading,
quartic, term are resolved by the well-known fact that the 4-point
disk-amplitude has total symmetry under permutations of the external
states (Green, Schwarz and Witten 1987).  One natural generalization
(Tseytlin 1997) is to evaluate  all higher-order  terms with the same
symmetrized-trace prescription, but 
this  fails to reproduce some
known facts about the open-string effective action
(Hashimoto and Taylor 1997, see also Brecher and Perry 1998, Brecher
1998).  Since commutators of field strengths cannot be  separated
in invariant fashion from higher-derivative (`acceleration') terms,
there is in fact no reason to expect that a  non-abelian
D-brane action  in a simple, closed form will  be found.

 Having summarized the known facts about  effective D-brane
 actions, we will spend the remainder of this review  justifying
 them, and exhibiting some of their  salient features.


\subsection{Type-I theory}
\label{subsec:typeI}

We first consider the special case $p=9$,
which  allows  contact  with the familiar action of the type-I 
string theory. The type-I background has 32 D9-branes plus  an
orientifold  (Sagnotti 1988, Ho\v rava 1989a), which  truncates
$U(32)$ to $SO(32)$  and 
projects out of the spectrum 
all antisymmetric-form fields other
 than  $C^{(2)}$ and its dual $C^{(6)}$. 
Since the gauge fields live on the D9-branes, their action should be
given entirely by $I_{{\rm D}9}$, after appropriate truncation of the
unnecessary fields. Note, in contrast, that  the 
purely-gravitational part  of the type-I,  tree-level 
lagrangian  has contributions  from three distinct diagrams~:
sphere, disk and real projective plane. The gauge-field independent
pieces  in  $I_{{\rm D}9}$ -- representing the contributions of the disk
-- cannot, therefore, be directly matched to
 the effective supergravity action.

Expanding out the Born-Infeld action  in powers of the field
strength, neglecting the (leading)  cosmological term which is
removed  by the orientifold projection, and using the total symmetry
of the 4-point function,  we find
\begin{equation}
I_{\rm BI} = T_{(9)}^I (\pi\alpha^\prime)^2\;
 \int d^{10}x \; e^{-\Phi}   \left[  \;
 {\rm tr}\; (F_{\mu\nu}F^{\mu\nu})  -  
 {(\pi\alpha^\prime)^2\over 12}\;
    {\rm tr}\; (t_8 F^4) +    \cdots \; \right]
\label{eq:expa}
\end{equation}
where  the dots stand for  higher orders  in
 $\alpha^\prime$, the  $F_{\mu\nu}$ are hermitean, 
and $t_8$ is the well-known eight-index tensor of string theory
(without its $\epsilon$ piece)  
\begin{equation}
\eqalign{
t_8  F^4 = \ & 
16\; F_{\mu\nu}F^{\nu\rho}F^{\lambda\mu}F_{\rho\lambda}
+8\; F_{\mu\nu}F^{\nu\rho}F_{\rho\lambda}F^{\lambda\mu}\cr
& -4\; F_{\mu\nu}F^{\mu\nu} F_{\rho\lambda}F^{\rho\lambda} 
-2\; F_{\mu\nu} F_{\rho\lambda}F^{\mu\nu}  F^{\rho\lambda} \ .
\cr}
\end{equation}
The quartic piece can be written alternatively as a 
symmetrized trace,
\begin{equation}
 {\rm tr}\;(t_8 F^4)  = 24 \; {\rm Str} \; \Bigl( F^4 - {1\over 4}
 (F^2)^2 \Bigr)\ ,
\end{equation}
with   matrix multiplication of the Lorentz indices implied.
Expanding out similarly the Wess-Zumino action of  the D9-branes, 
which  have of course a trivial  normal bundle, we find 
\begin{equation}
 I_{\rm GS}=   T_{(9)}^I\; (\pi\alpha^\prime)^2 \;  \int d^{10}x \;
  \left[\;  C^{(6)} \wedge {\rm X}_4 +\; (\pi\alpha^\prime)^2 \;
 C^{(2)} \wedge {\rm X}_8\; \right]\ , 
\end{equation}
where
\begin{equation}
\eqalign{
{\rm X}_4 &=   2\; {\rm tr}\; F^2
\; + \; \cdots  \cr
{\rm X}_8 &=   \;  {2\over 3}\; {\rm tr}\; F^4\; +\;
 {1\over 12}\; {\rm tr}\; F^2\; {\rm tr}\; {\cal R}^2\;
+ \cdots  \ .  \cr}
\end{equation}
Here the dots stand  for purely-gravitational corrections to 
${\rm X}_4$ and ${\rm X}_8$, multiplication is in the sense of forms, 
and  we recall that ${\cal L}_{\rm WZ}$ was 
defined only up to total derivatives.

  One can  recognize in the above expressions many of the standard
features of $SO(32)$ superstring theory.  The terms in $I_{\rm GS}$ are
the Green-Schwarz couplings that cancel the hexagon anomaly, 
as shown  in figure 7  (Green and Schwarz 1984, 1985a,b)
They have been  calculated directly from the disk
diagram by Callan {\em et al} (1988).
They are often expressed in
terms of traces (`Tr') in the adjoint representation of $SO(32)$, via
the relations
\begin{equation}
{\rm Tr}\; F^2 = 30\; {\rm tr}\; F^2 \ , \ \ \ {\rm Tr}\; F^4 =
 24\; {\rm tr}\; F^4 + 3\; ({\rm tr}\; F^2)^2 \ .
\end{equation}
This is 
less economical but more natural from the point of view of the
effective supergravity.  The  anomalous Green-Schwarz couplings are,
furthermore,
related, through space-time supersymmetry, to the two leading terms
which  we have exhibited in the
expansion of the Born-Infeld action 
(de Roo {\em et al} 1993,  Tseytlin 1996a,b,
 see also Lerche 1988).
Comparing the coefficients of the various  terms 
is a non-trivial check of our normalizations.
\footnote{I thank J. Conrad for discussions on this point.}
 For instance,  the
 tensor structure  mutliplying    ${\rm tr}\; F^4$ 
 has  the correct  supersymmetric form, 
$t_8 - {1\over 4} \epsilon_{10} C^{(2)}$ (Tseytlin 1996a) .
 The two terms in the ${\rm
  X}_8$ polynomial also have the right  relative weights
(Green {\em et al} 1987).  Finally, we can put the
quadratic Yang-Mills  lagrangian  in  the standard form,   
 $ {\rm tr} (F_{\mu\nu}F^{\mu\nu}) /2  g^2_{(10)}$, with  
\begin{equation}
{g^4_{(10)}\over  \kappa_{(10)}^2} = 2^{11} \pi^7 \alpha^{\prime\; 2} \ .
\end{equation}
This  is, indeed, the  relation  between the type-I gauge
 and gravitational  coupling constants
(Sakai and Abe 1988,  see also 
Shapiro and Thorn 1987,  Dai and Polchinski 1989 for the bosonic case).

\begin{figure}
%
%
\begin{center}
\leavevmode
\epsfxsize=10cm
\epsfbox{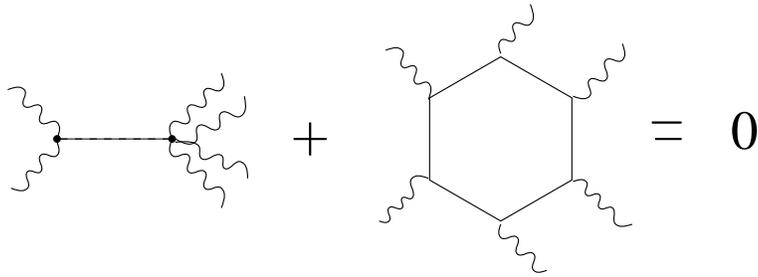}
\end{center}
\caption{  The Green-Schwarz
  anomaly-cancellation mechanism~:  the  classical interference of the
  two vertices in $I_{\rm GS}$, through the exchange of a  RR
  two-form,  cancels 
the one-loop hexagon anomaly.
\label{fig:GreenSchwarz}}
\end{figure}


\subsection{The power of T-duality}
\label{subsec:power}

   The effective D-brane actions (\ref{eq:Daction}--\ref{eq:WZ}) must
   be compatible with the T-duality transformations discussed in
   subsection 4.2. Indeed, being  discrete gauge symmetries, 
   T-dualities  leave the entire open and closed 
string theory, and hence also  its
   low-energy limit,  unchanged  (Dai
   {\em et al} 1989, Ho\v rava 1989b, Green 1991a). 
 Now  T-duality transforms in general a D$p$-brane to a
   D$p^\prime$-brane, with $p^\prime\not= p$, and exchanges
  transverse brane coordinates with  worldvolume gauge fields. Thus
 combined with  reparametrization invariance, 
it  puts stringent constraints on the
   dynamics of these latter
(Bachas 1996, Douglas 1995, Bergshoeff {\em et al} 1996,
Bergshoeff and de Roo 1996, Green {\em et al} 1996). 
 \footnote{Extended supersymmetry also relates brane coordinates
 to gauge fields, thereby constraining the dynamics of the latter.}

   We have already explained how  the simplest duality --
   inversion of the radius of a compactification circle -- maps
   wrapped  D$p$-branes  to transverse D$(p-1)$ branes, and
   vice versa.  It also maps the corresponding component of the
   $p$-dimensional  gauge field, to the extra  transverse coordinate of
   the lower brane, 
\begin{equation}
 A_{0,\cdots , p-1}^{\;\prime}  =  {A_{0,\cdots , p-1}}\ ,  \ \ \ \ 
 {Y_p}^{\prime}  =  2\pi\alpha^\prime\; A_p \ , \ \ \ \  
 Y_{p+1,\cdots,9}^\prime    =    Y_{p+1,\cdots,9}\ . 
\end{equation}
Here the D$p$-brane spans the dimensions $(1,\cdots,p)$,
$x^p = x^p+ 2\pi R_p$ is the spatial coordinate that we dualize,
and both branes are being parametrized in static gauge.
All fields are functions of the common worlvolume 
coordinates $\zeta^{0,\cdots,p-1}$.
In addition, the D$p$-brane fields may depend on the coordinate $\zeta^p$,
or equivalently on its  conjugate momentum, in which case the
D$(p-1)$-brane fields depend on the dual winding  in the $p$th
dimension. This  dependence drops  out 
in the limit  $R_p =\alpha^\prime/R_p^\prime\to 0$, since
momentum (or  winding prime)  modes become infinitely massive
and decouple. Thus a non-compact transverse D-brane coordinate can be
thought of as a gauge field in some vanishingly-small internal dimension.

 Let us focus attention to  the case $p=1$. The effective action of a
type-IIB D-string, wrapping around a tiny (first) dimension reads
\begin{equation}
I   =  T_{(1)} \int d^2\zeta \;
\Bigl[  \sqrt{1 -
 (\partial_0 Y^m)^2- ( 2\pi\alpha^\prime\; F_{01})^2}\;
+\; {\widehat C}_{01} +  2\pi\alpha^\prime  F_{01}\;{\widehat C}
 \Bigr]\ .
\label{eq:Dstr}
\end{equation}
We have here assumed for simplicity a
 flat space-time and  vanishing $\Phi$ and
 $B_{\mu\nu}$  backgrounds. The   T-duality transformation $R_1^{\;\prime}
 = \alpha^\prime/R_1$  changes  the D-string to a D-particle,
 the worldvolume electric field $F_{01} = \partial_0 A_1$
to a velocity,  and the RR backgrounds as follows~:
\begin{equation}
C_1^{\;\prime}  =  C \ ,\  \ \  C_{\mu}^{\;\prime} =  C_{\mu 1} \
\ \ (\mu\not= 1) .
\end{equation}
The effective D-string action transforms therefore to
\begin{equation}
I^\prime = 
 T_{(0)}^\prime
 \int d\zeta^0\;  \Bigl[  \sqrt{1- ( \partial_0 {\vec  Y^{ \;\prime}}   )^2
}\ 
+ \  {\widehat C}_{0}^{\; \prime}   \Bigr]  \ \ .
\label{eq:Dpart}
\end{equation}
This is indeed   the effective action of a  D-particle  
coupling  to the dual  RR one-form  background, as should be expected. 

   If one neglects acceleration terms, the action of a charged point
 particle in flat spacetime is  fixed uniquely  by Lorentz
 invariance, together with  invariance under reparametrizations 
of the worldline. Running the logic backwards, we could  thus start with
 the D-particle action (\ref{eq:Dpart})
and invoke  T-duality plus locality to determine 
the (abelianized) gauge dynamics on the worldvolume of 
D-strings,   or  higher D-branes. This is quite remarkable, since
 from simple kinematic constraints  one is  deriving apparently
 non-trivial information about  open-string  gauge dynamics. 
The anomaly-cancelling Green-Schwarz terms are,  for example,
 related to the covariantization of Polchinski's coupling 
 (Douglas 1995), while the speed of light ($c=1$) is mapped, under T-duality,
 to the limiting electric field of the Born-Infeld action (Bachas
 1996).  Notice that,  due to  Regge behaviour, 
$c$  appears  as a dynamical rather than purely kinematic  limit.  
The corresponding  dissipation mechanism is 
 the pair production calculated  in subsection 5.3.


\section{Topological aspects of brane  dynamics}

   The effective gauge theories   on the worldvolume of  D-branes 
 have  a rich spectrum  of both perturbative and non-perturbative
excitations. These are worldvolume projections of the various branes
from  the bulk which,
 like fundamental strings,  can terminate on the D-branes of
interest, or form with them   stable bound states.
 Much can be learned about these
dynamics  by simple topological considerations of the
worldvolume fluxes and charges,  and  of their spacetime counterparts. 
We  conclude this  guided tour of D-branes with a brief discussion of
such  issues.

\subsection{Branes inside  branes}
\label{subsec:impl}

 One  immediate consequence  of the Wess-Zumino 
action (\ref{eq:WZ})
is that the worldvolume  gauge fields and the geometry
 can induce different 
RR  charges  on D-branes.  We will illustrate this phenomenon
with some  concrete examples of D$p$-branes wrapped around a compact
cycle $\Sigma_k$, such as a $k$-torus  or a supersymmetric 
$k$-cycle of a Calabi-Yau  space. To simplify the discussion we assume
that the   target space is 
a direct product of d-dimensional Minkowski space ($\RR^d$)
times a  compactification manifold ($ \Sigma_{10-d}$),
 and that the  brane worldvolume can be  factorized accordingly,
 ${\cal W}_{p+1} = \Sigma_k \times {\cal W}_{p-k+1}$  
with   $\Sigma_k  \subset
 \Sigma_{10-d}$ and ${\cal W}_{p-k+1}\subset \RR^{d}$. 
We also  assume 
antisymmetric-tensor backgrounds that are covariantly constant on the
compactification manifold, as well as a vanishing dilaton field.

 Our first example is a  D$2$-brane  wrapped around
a two-cycle  $\Sigma_2$,  on which we turn on a 
(quantized) magnetic flux,   
\begin{equation}
{1\over 2\pi}  \int_{\Sigma_2}  \; F\; =\;   k\ \   .   
\end{equation}
This gives rise to a 
 Wess-Zumino coupling
\begin{equation}
I_{\rm WZ}\; =\;
 T_{(2)} \int_{{\cal W}_{3}}
{\widehat C}^{(3)}\; +\; k\; T_{(0)}
\int_{{\cal W}_{1}} {\widehat C}^{(1)}  \ , 
\end{equation}
showing  that the D2-brane 
has  acquired   $k$ units of RR one-form charge.
We are therefore describing a 
 configuration of a  D2-brane bound to 
 $k$ type-IIA D-particles,
 or equivalently of a (transverse) membrane boosted
along the hidden eleventh dimension of M-theory.

 To confirm this latter
interpretation, notice that the Dirac-Born-Infeld action in 2+1
dimensions reads
\begin{equation}
 I_{\rm DBI} = T_{(2)}\;\int d^3\zeta\; 
 \sqrt{- {\widehat G}}\;\sqrt{
 1 + 2\pi^2\alpha^{\prime\ 2}
 F_{\alpha\beta} F^{\alpha\beta} }\ ,
\end{equation}
with indices  raised by  the induced metric.
Extremizing this action for a 
 static D2-brane yields a  magnetic field proportional
to the volume form ($\omega_{\Sigma_2}$) of the 2-cycle, 
\begin{equation}
 F  =  2\pi k\; { \omega_{\Sigma_2} \over 
 \int\omega_{\Sigma_2} }  \ .
\label{eq:uniform} 
\end{equation}
Using the relations between type-IIA and M-theory  scales  we can write
the energy of this configuration (for zero RR backgrounds)  as 
\begin{equation}
{\cal E} = 
 \sqrt{ M^2  +
 \left({k\over R_{11}}\right)^2}\   \ \ \ {\rm with}\ \ \ 
M=  T_{(2)}\; \int\omega_{\Sigma_2}\ .
\end{equation}
This is the expected energy of an excitation of  mass $M$, 
 carrying   $k$ units of momentum in the eleventh
dimension. The scalar  dual to the vector field on the worldvolume
 can  be in fact interpreted as the eleventh coordinate
of the membrane  (Townsend 1996a, Schmidhuber 1996),
\begin{equation}
{2\pi\alpha^\prime F_{\alpha\beta} \over 
 (1+ 2\pi^2\alpha^{\prime\ 2} F^2)^{1/2}} \;  = \; \sqrt{-{\widehat G} }\;
 \epsilon_{\alpha\beta\gamma}\; \partial^\gamma Y^{11}\ . 
\end{equation}
This duality transformation maps indeed
the magnetic field  (\ref{eq:uniform}) 
to a  uniform motion along $x^{11}$, as should be the case.

 Our  second example consists of  $n$ coincident D4-branes
 wrapping around a  four-cycle $\Sigma_4$. A non-abelian  $k$-instanton
 configuration,
\begin{equation}
-{1\over 8\pi^2}
 \int_{\Sigma_4} {\rm tr}\;  F\wedge F =  k \ \ ,
\end{equation}
induces  RR one-form charge equal  to that carried by
 \begin{equation}
  k \;+ \; {n\over 48} \; \Bigl( p_1({\cal T}) -  p_1({\cal
 N})  \Bigr)
\end{equation}
(anti)D-particles.  This is a very fruitful interpretation, which
allows an identification of certain 
 multi-instanton moduli spaces  with  vacuum manifolds 
 of  appropriate supersymmetric
gauge models  (Witten 1996b,
Bershadsky {\em et al} 1996b, Vafa 1996, Douglas 1995,
Douglas and Moore 1996). The dimension of such a moduli space 
enters, in particular, in the simplest microscopic derivation of the
Bekenstein-Hawking entropy from  D-branes (Strominger and Vafa 1996,
Callan and Maldacena 1996).

 Consider, for example, the case of $\Sigma_4$ a  four-torus,
so that our  configuration consists of   $n$ flat
D4-branes and $k$ (anti)D-particles. As explained  in subsection 5.1,
such a configuration leaves  eight unbroken  supersymmetries, and has
the following content of low-lying open-string excitations: 
(i) a ten-dimensional $U(n)$ vector multiplet reduced down to the
worldvolume of the four-branes, (ii) a ten dimensional $U(k)$ vector
multiplet reduced likewise to  the worldline of the particles, and
(iii) a six-dimensional hypermultiplet, in the $(n, {k})$
representation of the gauge group, and living  also on the particle
worldline. In terms of the unbroken supersymmetries, the adjoint
fields decompose into vector plus hypermultiplets.
The vacuum manifold of this effective theory has a Coulomb
branch along which the gauge group breaks generically 
to $U(1)^n\times U(1)^k$,
and a Higgs branch along which only a single  U(1) remains unbroken.
These correspond, respectively, to  D-particles separated from the
D4-branes in the transverse space, or bound to them as
finite-size instantons. The dimension around a generic point on the
Higgs branch is  given by the total number of scalar fields in the
hypermultiplets, 
 minus the number of   gauge transformations and  of
D-flatness conditions
(see for example Hassan and Wadia 1997)
\begin{equation}
{\rm dim} {\cal M}^{n}_{k} =
 4(nk + n^2 + k^2) - (n^2+ k^2-1)-
3(n^2+ k^2-1)  \; =\;  4(nk  +1)\ .
\end{equation}
This is indeed the dimension of the moduli space of $k$ U(n)
instantons on a flat torus. Similar constructions work for instantons
on a K3 surface, for which the first Pontryagin class $p_1(K3) = -48$
( Bershadsky {\em et al} 1996b, Vafa 1996),
as well as for instantons on a  ALE space (Douglas and Moore 1996).

  Our last example is a type-IIB D-string  winding around a stable cycle
$\Sigma_1$ of unit radius. Turning on a worldsheet electric field
gives a coupling linear in
the Neveu-Schwarz antisymmetric tensor,
\begin{equation}
I_{{\rm D}1}\;   =\; {1\over 2\pi\alpha^\prime} \int d^2\zeta\; 
{ \Pi}_1 \;
{\widehat  B}_{01} + \cdots
\label{eq:fstr} 
\end{equation}
where  ${ \Pi}_1 = \delta{\cal L}/\delta \partial_0A_1$ is the momentum
conjugate to $A_1$. We have here gone to the $A_0=0$ gauge,
 used $\zeta^1$
to parametrize the stable cycle, and set for simplicity the RR
backgrounds to zero. Since the Wilson line  $\oint d\zeta^1 A_1 $ is a
periodic variable with period $2\pi$, its conjugate momentum in the
quantum theory is integer,  $\oint {d\zeta^1} \Pi_1 = 2\pi  q$.
The coupling (\ref{eq:fstr}) describes then precisely  the gauge charge 
carried by $q$ 
fundamental winding strings,  bound to
the D-string under consideration
(Witten 1996a, Callan and Klebanov 1996, Schmidhuber 1996).
This is in accordance with the prediction of SL(2,Z)
duality, which requires the existence of  subthreshold bound states of
$p$ D-strings and $q$ fundamental strings, for all  pairs $(p,q)$ of
relatively prime integers (Schwarz 1995). 

\vskip 0.9cm

\subsection{Branes ending on branes}
\label{subsec:ending}

   The coupling of   D-branes to $B_{\mu\nu}$  can be 
understood from a simple worldsheet argument. Under a gauge
transformation $\delta B = d\Xi$, with  $\Xi$   an arbitrary  one-form,
the action of a fundamental string changes by a boundary term
\begin{equation}
\delta_{\Xi} I_{F} = {1\over 2\pi\alpha^\prime}
 \oint_{\partial\Sigma} d\xi^a\; {\widehat \Xi}_a \ .
\end{equation}
To cancel this variation, we must assume that the gauge fields living
on the worldvolumes of D-branes have also a universal transformation
$\delta A_\alpha = -
{\widehat \Xi}_\alpha/ { 2\pi\alpha^\prime}$.
  This explains the
appearance of the gauge-invariant combination 
${\hat B} + 2\pi\alpha^\prime F$  in the Dirac-Born-Infeld
action. Invariance of the Wess-Zumino action, on the other hand,
requires that the (sum over all) RR potentials transform as 
\begin{equation}
\delta_{\Xi} C = C\wedge e^{d\Xi} \ .
\end{equation}
The RR antisymmetric tensors  have of course  their own independent gauge
transformations,
\begin{equation}
\delta_{\Lambda} C = d\Lambda \ , 
\label{eq:simple}
\end{equation}
with $\Lambda$ (a sum of) arbitrary forms. 
Redefining  the  RR potentials, ${\tilde C} \equiv
{ C}\wedge e^{-B}$, so as to make them  invariant under the
$\Xi$-transformations, modifies the $\Lambda$-transformations which
would in this case  mix the RR forms and  the Neveu-Schwarz tensor.
\footnote{I thank E. Kiritsis for this argument.}

  This argument confirms what we have used from the very beginning, i.e.
 that a fundamental string may  end on any
D-brane, on whose worldvolume it couples  as an elementary electric charge.
The dynamics  of such open strings can in fact be analyzed  from the
 viewpoint of the worldvolume 
Born-Infeld action (Callan and Maldacena 1998, Gibbons 1998).
We can, however, also generalize the argument to see what other branes
 can end on D-branes (Strominger 1996).
Consider indeed the variation of the Wess-Zumino action of a D$p$-brane
under a gauge transformation of the RR $(p+1)$-form,
\begin{equation}
\delta_{\Lambda} I_{Dp}\; = \;  T_{(p)}
 \int_{\partial {\cal W}_{p+1}} {\hat \Lambda}^{(p)}\ ,
\label{eq:varn}
\end{equation}
where $\partial W_{p+1}$ is the boundary of the brane worldvolume.
We may  cancel this variation if the boundary happens to lie on the
 worldvolume of a
D$(p+2)$-brane, on which it appears as the trajectory of a
$(p-1)$-dimensional magnetic charge~! Indeed,  the anomalous
Bianchi identity on the worldvolume of the D$(p+2)$-brane reads 
\begin{equation}
d\wedge F =  2\pi  \delta^{(3)} (\partial {\cal W}_{p+1}) \ ,  
\end{equation}
where we have used the normalization of electric charge to one. It can
be checked that the 
variation  of the Wess-Zumino action of the higher brane will then
cancel precisely (\ref{eq:varn}). 

  We thus conclude that 
 D-strings can  terminate
on D3-branes,  on whose worldvolume they appear as magnetic monopoles,
that D2-branes can terminate  on D4-branes, and so on for
higher $p$. Notice that if the branes are orthogonal, 
the number of dimensions along  one or other of the branes
 but not both   is
exactly four. These configurations leave therefore one quarter  of unbroken
supersymmetries. Various duality transformations
 map the above examples to other configurations of branes. Note in
 particular that lifted to M-theory, the D2-D4 brane configuration
 teaches us that the M-theory membrane can terminate on a M-theory
 fivebrane.

\subsection{Branes created by crossing branes}
\label{subsec:impl}

    The final point I want to discuss has to do with the role of
Wess-Zumino terms in cancelling chiral anomalies.
We already saw this  in the context of type I theory, but the
phenomenon is more general  and can in fact be used to fix completely
the form of the 
Wess-Zumino couplings (Green {\em et al} 1997).  Consider 
for example two  stacks  of $n$ and $n^\prime$
D5-branes, spanning worldvolumes
 ${\cal W}_6$ and
${\cal W}_6^{\;\prime}$, which have generically a two-dimensional
intersection, ${\cal I} = {\cal W}_6 \cap  {\cal W}_6^{\;\prime}$.
Let us  concentrate on the gauge part of the Wess-Zumino action of the
first stack. Under a worldvolume gauge transformation, this has an anomalous
variation given by
\begin{equation}
\delta_{\xi}  I_{D5}  = T_{(5)} (2\pi\alpha^\prime)^2
 \int_{{\cal W}_6} d {\hat H}^{(3)} \wedge {\rm tr
  } (\xi F) =  {n^\prime \over 2\pi} 
 \int_{\cal I} {\rm tr
  } (\xi  F) \ \ ,
\label{eq:anomalous}
\end{equation}
with $\xi$ in the Lie algebra of $U(n)$. 
We have here used the  standard descent formulae
\begin{equation}
{\rm tr}( F \wedge F)  = d \omega_3(A)\ \ \ {\rm and}
 \ \ \delta_{\xi} \omega_3(A) = d\; {\rm tr
  } (\xi  F)\ , 
\end{equation}
with  $\omega_3(A)$  the Chern-Simons three form, as well as the 
 (anomalous)
 Bianchi identity 
\begin{equation}
 d {\hat H}^{(3)} = 2\kappa_{(10)}^2 T_{(5)}\;
 n^\prime \delta^{(2)}({\cal I}  )\ .
\end{equation}
This is the projection on  ${\cal W}_6$ of the bulk Bianchi identity
showing that  the prime  branes are magnetic
 sources for the RR two-form.

 Thus gauge invariance
seems to be violated on the  intersection, but the anomaly can be
precisely cancelled by $n^\prime$ chiral fermions in the fundamental
representation of $U(n)$.  But as we have already explained  in
section 5.1, string theory provides us precisely with 
the required fermions -- the massless states of the fundamental
strings stretching from one to the other stack of D5-branes, and
transforming in the $(n, {\bar n^\prime})$ representation of
the $U(n)\times U(n^\prime)$ gauge group. Reversing the argument,
since the embedding theory is non-anomalous, the presence of the
Wess-Zumino couplings is necessary to cancel the apparent 
violation of charge
conservation on the intersection, by inflow from the bulk of the D5-branes. 
 The gravitational anomaly of the intersection fermions cancels
similarly the anomalous variation of the gravitational Wess-Zumino
action.  
To see how one must use the  (Whitney sum) decompositions of the
tangent bundles (since the branes fill all dimensions)
\begin{equation}
{\cal T}_{{\cal W}_6} = {\cal T}_{\cal I}
 \oplus {\cal N}_{{{\cal W}_6}^\prime},
\end{equation}
and similarly for the prime  brane, together with the multiplicative
property of the roof genus. It follows that the anomalous variations
of the pullback normal bundles  cancel between the
two stacks of D5-branes, while those of the bundle tangent
 to the intersection add up and cancel
against the anomalous fermion loop.

\begin{figure}
%
%
\begin{center}
\leavevmode
\epsfxsize=11.5cm
\epsfbox{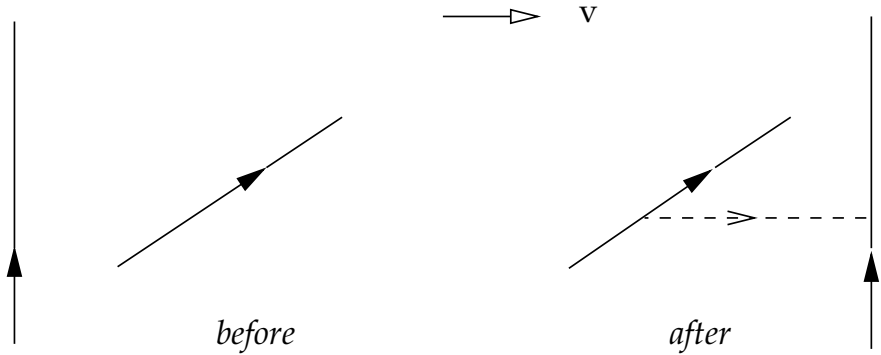}
\end{center}
\vskip0.3cm
\caption{  Anomalous creation of a stretched string
  when two orthogonal D4-branes cross. 
\label{fig:branecr}}
\end{figure}

 The anomalous inflow of charge,  required to cancel the chiral
 anomaly on the intersection,
 has an interesting T-dual interpretation 
(Bachas {\em et al} 1997, 
Danielsson {\em et al} 1997, Bergman {\em et al} 1998).
Consider indeed two (stacks of) D4-branes oriented so as to have a
unique common transverse dimension, say $x_1$. The lowest-lying state
of an open string stretching between the two stacks is a chiral
 fermion, but since it is completely localized in space the role of
 momentum is played by its (oriented) stretching. It thus satisfies
 the T-dual Weyl equation,
\begin{equation}
p_0 = T_F \delta x_1\ ,
\end{equation}
 with $\delta x_1$ the transverse displacement of the two D4-branes.  
Now as the D4-branes move continuously past each other, the above
energy level crosses continuously the zero axis. Thus in the
second-quantized theory a string must be 
 anomalously created or destroyed, as illustrated in figure 8.

Since this is a topological phenomenon, we should expect it to commute
with any (sequence of)  dualiy transformations. Consider in particular
the following chain,  
 \[
\begin{array}{ccc}
({\rm IIA})\ \ \ \ \  &\ \ {\rm D}(2345) \bigotimes
  {\rm D}(6789)  \hookrightarrow
{\rm F}(1) \\
 \ & \\
 \  & \Biggl\downarrow { T{(6)\ \ \ } }   \\
 \ & \\
({\rm IIB})\ \ \ \ \   & {\rm D}(23456) \bigotimes {\rm D}(789)
\hookrightarrow {\rm F}(1) \\
  \ & \\
\  & \Biggl\downarrow { S{\ \ \ \ \ \ } } \\
\ & \\
({\rm IIB})\ \ \ \ \  & {\rm NS}(23456) \bigotimes {\rm D}(789)
\hookrightarrow  D(1) \\
\ & \\
 \  & \Biggl\downarrow { T(56) }  \\
 \ & \\
({\rm IIB})\ \ \ \ \   & \ \ \ \ \ \ {\rm NS}(23456) \bigotimes {\rm D}(56789)
\hookrightarrow {\rm D}(156)
\end{array}
\]
\vskip 0.2cm
\noindent
Here $F$, $D$ and $NS$ denote a fundamental string, a D-brane and a
Neveu-Schwarz fivebrane, the dimensions which these branes span  are
indicated inside parentheses, and $X\otimes Y \hookrightarrow Z$
 means `$Z$
is being created when $X$ crosses $Y$'.
The sequence of T- and S-duality transformations tells us that 
a D3-brane must be created when a NS-brane and a D5-brane, sharing two common
dimensions, cross  each other. From the fermionic character of the
original stretched fundamental string, we also know that only a single
streched brane in a supersymmetric state is allowed (Bachas {\em et
  al} 1998).  These two basic rules of brane  engineering 
have been  indeed  postulated by Hanany and Witten (1997), in
order to avoid immediate contradictions with the known behaviour of
three-dimensional supersymmetric gauge models.

  I stop here because  time is up -- not because the subject has been
exhausted. The reader has hopefully acquired the tools, as well as the
motivation, to move on to some of the exciting applications of
D-branes to gauge theories  and black-hole physics.

\section*{Acknowledgments}
 I thank David Olive, Pierre van Baal and Peter West for organizing a
wonderful  Newton-Institute workshop.
 These lectures were also
presented at the 1997 Trieste Spring School on `String Theory, Gauge
Theory and Quantum Gravity', and in a shorter format at the `31st
International Symposium Ahrenshoop' in Buckow. 
I thank the organizers for the invitations  to speak, and present to them
my sincere apologies for failing to meet their (generously-elastic) 
proceedings deadlines.
I am finally indebted to P. Bain, J. Conrad,  E. Cremmer,   G. Gibbons,
 M.B. Green, A. Greenspoon,  M. Henneaux, 
A. Kehagias, E. Kiritsis, 
J.X. Lu, S. Mukhi,  H. Partouche, B. Pioline, A. Polychronakos,  S. Sethi
and  P. Vanhove for comments and discussions that
 helped improve this manuscript.

\section*{Note on conventions}

Throughout the text I have  used  $X^\mu(\xi^a)$
for the coordinates of a fundamental string, $Y^\mu(\zeta^\alpha)$
for those of a D-brane, and $x^\mu$ for the space-time coordinates.
I reserve the capital N to count supersymmetries, and the 
little $n$ for the
number of D-branes.
Hats denote pullbacks of supergravity fields from the bulk onto the
worldvolumes  of branes.  $T_F=
1/2\pi\alpha^\prime$ is the tension of the fundamental string, 
not to be confused with the worldsheet
supercurrent which I have denoted $J_F$. 
I use `{\em et al}'  when referring to papers
with three or more coauthors, and indicate in parenthesis the
publication year when available, or the year of submission to the
archives otherwise. All authors and all archive numbers can be found
in the bibliography at the end.

\section*{References}



\refce
Abou Zeid, M., Hull, C.M. (1997) `Intrinsic Geometry of D-Branes',
{\em Phys. Lett. \bf B404}, 264--270, hep-th/9704021.

\refce
Aganagic, M., Popescu, C., Schwarz, J. (1997a)
`D-Brane Actions with Local Kappa Symmetry', {\em Phys. Lett. \bf B393},
311--315, hep-th/9610249.

\refce
Aganagic, M., Popescu, C., Schwarz, J. (1997b) `Gauge-Invariant
 and Gauge-Fixed D-Brane Actions', {\em Nucl. Phys. \bf B495}, 99--126,
 hep-th/9612080.

\refce
Alvarez, E., Barbon, J.L.F., Borlaf, J. (1996) `T-duality for open strings',
{\em Nucl. Phys. \bf B479}, 218--242, 
hep-th/9603089. 

\refce
Alvarez, O. (1981)  `Static Potential in String Models',
 {\em Phys. Rev. \bf D24}, 440--449.

\refce
Alvarez-Gaume, L., Zamora, F. (1997) `Duality in Quantum Field Theory
(and String Theory)', lectures at `The Workshop on Fundamental Particles and
Interactions', Vanderbilt University,
and CERN-La Plata-Santiago de Compostela School of Physics, 
hep-th/9709180.

\refce
Amati, D., Ciafaloni, M., Veneziano, G. (1987) `Superstring Collisions
at Planckian Energies',
{\em Phys. Lett. \bf B197}, 81--88.

\refce
Amati, D., Ciafaloni, M., Veneziano, G. (1989) `Can Spacetime be
Probed below the String Size~?',
{\em Phys. Lett. \bf B216}, 41--47.

\refce
Andreev, O.D., Tseytlin, A.A. (1988) `Partition Function
Representation for the Open Superstring Effective Action',
{\em Nucl. Phys. \bf B311}, 205--252. 

\refce
Antoniadis, I., Partouche, H., Taylor, T.R. (1998)
`Lectures on Heterotic-Type I Duality', {\em
Nucl. Phys. Proc. Suppl. \bf
61A}, 58--71, hep-th/9706211.

\refce
Arvis, J.F. (1983) `Deux Aspects des Th{\'e}ories de Cordes Duales:
la Th{\'e}orie de Liouville Supersym{\'e}trique et le Potentiel $q\bar q$
dans le Mod{\`e}le de Nambu',
 PhD thesis, University of Paris. 

\refce
Aspinwall, P. (1996) `K3 Surfaces and String Duality', 
lectures at  TASI 96, hep-th/9611137.

\refce
Bachas, C., Porrati, M. (1992) `Pair Creation of Open Strings in an
Electric Field', {\em Phys. Lett. \bf B296}, 77--84, 
hep-th/9209032. 

\refce
Bachas, C. (1996)  `D-brane Dynamics',
{\em Phys. Lett. \bf B374}, 37--42, hep-th/9511043.

\refce
Bachas, C., Kiritsis, E. (1997) `$F^4$ Terms in N=4 String Vacua', 
{\em Nucl. Phys. Proc. Suppl. \bf 55B}, 194--199, 
 hep-th/9611205. 

\refce
Bachas, C. (1997a) `(Half) a Lecture on D-branes',
 Proceedings of the Workshop on `Gauge Theories, Applied
Supersymmetry and Quantum Gravity', Sevrin A. {\em et al} eds., 
Imperial College Press, pp. 3--22,  hep-th/9701019. 

\refce
Bachas, C. (1997b) `Heterotic versus Type I', talk at `Strings 97', 
hep-th/9710102. 

\refce
Bachas, C., Douglas, M.R., Green, M.B. (1997)
 `Anomalous Creation of Branes', {\em JHEP \bf 07},002, 
hep-th/9705074. 

\refce
Bachas, C.,  Green, M.B., Schwimmer, A.  (1998)
`(8,0) Quantum mechanics and symmetry enhancement in type
 I' superstrings', {\em JHEP \bf 01}, 006, hep-th/9712086.

\refce
Balasubramanian, V., Larsen, F. (1997) `Relativistic Brane
Scattering', {\em Nucl. Phys. \bf B506}, 61--83,
 hep-th/9703039. 

\refce
Banks, T, Susskind, L. (1995) `Brane - Anti-Brane Forces', 
hep-th/9511194.

\refce
Banks, T., Douglas, M.R., Seiberg, N. (1996) `Probing F-theory With Branes',
{\em Phys. Lett. \bf  B387}, 278--281, hep-th/9605199.

\refce
Banks, T.,  Fischler, W.,  Shenker, S.H.,  Susskind, L. (1997)
`M Theory As A Matrix Model: A Conjecture', {\em Phys. Rev. \bf D55},
5112--5128, hep-th/9610043. 

\refce
Banks, T., Seiberg, N., Silverstein, E. (1997) `Zero and
One-dimensional Probes with N=8 Supersymmetry', {\em Nucl. Phys. \bf
B401}, 30--37,  hep-th/9703052.

\refce
Banks, T. (1998) `Matrix Theory', {\em Nucl. Phys. Proc. Suppl. \bf
  B67}, 180--224,  hep-th/9710231.

\refce
Bautier, K., Deser, S., Henneaux, M., Seminara, D. (1997)
`No Cosmological D=11 Supergravity', {\em Phys. Lett. \bf B406}, 49--53,
hep-th/9704131. 

\refce
Becker, K., Becker, M. (1997) `A Two Loop Test of Matrix Theory',
{\em Nucl. Phys. \bf B506}, 48--60, hep-th/9705091.

\refce
Berenstein, D., Corrado, D.R., Fischler, W., Paban, S., Rozali,
M. (1996) `Virtual D-Branes', {\em Phys. Lett. \bf B384}, 93--97, 
hep-th/9605168.

\refce
Bergman, O., Gaberdiel, M.R., Lifschytz, G. (1998)
`Branes, Orientifolds and the Creation of Elementary Strings', 
{\em  Nucl. Phys. \bf B509} 194--215, hep-th/9705130.

\refce
Bergshoeff, E.,  Sezgin, E., Townsend, P.K. (1987) `Supermembranes and
Eleven-dimensional Supergravity', 
{\em Phys. Lett. \bf  B189 }, 75--78. 

\refce
Bergshoeff, E., de Roo, M.,
 Green, M.B., Papadopoulos, G., Townsend, P.K. (1996)
`Duality of Type II 7-branes and 8-branes',
{\em Nucl. Phys. \bf B470}, 113--135, 
 hep-th/9601150.

\refce
Bergshoeff, E., de Roo, M. (1996) `D-branes and T-duality', 
{\em Phys. Lett. \bf B380}, 265--272, 
 hep-th/9603123. 

\refce
Bergshoeff, E., Townsend, P.K. (1997) `Super D-branes', {\em
  Nucl. Phys. \bf B490}, 145--162, hep-th/9611173.

\refce
 Bergshoeff, E.,  Lozano, Y.,  Ortin, T. (1997) `Massive Branes',
hep-th/9712115.

\refce
Berkooz, M., Douglas, M.R., Leigh, R.G. (1996)
`Branes Intersecting at Angles', {\em Nucl. Phys. \bf B480}, 265--278,
hep-th/9606139.

\refce
Bershadsky, M., Sadov, V., Vafa, C. (1996a)
`D-Strings on D-Manifolds', {\em Nucl. Phys. \bf B463}, 398--414, 
hep-th/9510225.

\refce
Bershadsky, M., Sadov, V., Vafa, C. (1996b) `D-Branes and Topological
Field Theories', {\em Nucl. Phys. \bf B463}, 420--434, 
hep-th/9511222. 

\refce
Bianchi, M. (1997) `Open Strings and Dualities', 
Talk  at the V Korean-Italian Meeting on Relativistic Astrophysics,
hep-th/9712020.

\refce
Bigatti, D., Susskind, L. (1997) `Review of Matrix Theory', lectures
at Cargese 97,
hep-th/9712072.

\refce
Bilal, A. (1996) `Duality in N=2 SUSY SU(2) Yang-Mills Theory:
 A pedagogical introduction to the work of
 Seiberg and Witten', hep-th/9601007.

\refce
Bilal, A. (1997) `M(atrix) Theory : a Pedagogical Introduction',
hep-th/9710136.

\refce
Billo, M., Cangemi, D., Di Vecchia, P. (1997)
`Boundary states for moving D-branes', {\em Phys. Lett. \bf B400}, 63--70,
hep-th/9701190. 

\refce
Brax, P., Mourad, J. (1997) `Open supermembranes in eleven
dimensions', 
{\em Phys, Let. \bf B408}, 142--150,
hep-th/9704165. 

\refce
Brax, P., Mourad, J. (1998) `Open Supermembranes Coupled to M-Theory
Five-Branes', {\em Phys. Lett. \bf B416}, 295--302, 
hep-th/9707246. 

\refce
Brecher, D., Perry, M.J. (1998) `Bound States of D-Branes
 and the Non-Abelian Born-Infeld Action', 
hep-th/9801127. 

\refce
Brecher, D. (1998) `BPS States of the Non-Abelian Born-Infeld Action',
hep-th/9804180. 

\refce
Callan, C.G., Lovelace, C., Nappi, C.R., Yost, S.A. (1988)
`Loop Corrections to Superstring Equations of Motion',
 {\em Nucl. Phys. \bf B308}, 221--284. 

\refce
Callan, C.G., Harvey, J.H., Strominger, A. (1991a) `Worldsheet Approach
to Heterotic Instantons and Solitons', {\em Nucl. Phys. \bf B359}, 611--634.

\refce
Callan, C.G., Harvey, J.H., Strominger, A. (1991b) `Worldbrane Actions
for String Solitons', {\em Nucl. Phys. \bf B367}, 60--82.

\refce
Callan, C.G., Klebanov, I. (1996) `D-Brane Boundary State Dynamics',
{\em Nucl. Phys. \bf B465}, 473--486, hep-th/9511173. 

\refce
Callan, C.G., Maldacena, J. (1996) `D-brane Approach to Black Hole
 Quantum Mechanics', {\em Nucl. Phys. \bf B472}, 591--610,
hep-th/9602043. 

\refce
Callan, C.G., Maldacena, J. (1998) `Brane Dynamics
 from the Born-Infeld Action', {\em Nucl. Phys. \bf B513}, 198--212,
hep-th/9708147.

\refce
Cederwall, M., von Gussich, A., Nilsson, B., Sundell, P., Westerberg,
A. (1997a) ` The Dirichlet Super-Three-Brane
 in Ten-Dimensional Type IIB Supergravity', 
{\em Nucl. Phys. \bf B490}, 163--178, 
hep-th/9610148.

\refce
Cederwall, M., von Gussich, A., Nilsson, B., Sundell, P., Westerberg,
A. (1997b) `The Dirichlet Super-p-Branes in Ten-Dimensional
 Type IIA and IIB Supergravity', {\em Nucl. Phys. \bf B490}, 179--201, 
hep-th/9611159. 

\refce
Cederwall, M. (1997) `Aspects of D-brane actions', {\em Nucl.
Phys. Proc. Suppl. \bf B56}, 61--69, 
hep-th/9612153.

\refce
Cheung, Y-K.E., Yin, Z. (1997) `Anomalies, Branes, and Currents', 
hep-th/9710206.

\refce
Coleman, S. (1981) `The Magnetic Monopole Fifty Years Later', in
`Gauge Theories in High Energy Physics',
 Les Houches proceedings. 

\refce
Connes, A. (1994) `Noncommutative Geometry', Academic Press.

\refce
Conrad, J.O. (1997) `Brane
 Tensions and Coupling Constants from within M-Theory', 
 hep-th/9708031.

\refce
Cremmer, E., Julia, B., Scherck, J. (1978) `Supergravity Theory in
Eleven Dimensions', {\em Phys. Lett. \bf B76}, 409--412.

\refce
 Dai, J.,  Leigh,  R.G., Polchinski, J. (1989)
`New Connections Between String Theories',
  {\em  Mod. Phys. Lett. \bf A4}, 2073--2083.

\refce
Danielsson, U.H., Ferretti, G., Sundborg, B. (1996)
`D-particle Dynamics and Bound States', {\em Int. Jour. Mod. Phys. \bf
  A11}, 5463-5478, 
hep-th/9603081.  

\refce
Danielsson, U.H., Ferretti, G. (1997) `The Heterotic Life of the
D-particle',  {\em Int. J. Mod. Phys. \bf A12}, 4581--4596, 
hep-th/9610082.

\refce
Danielsson, U.H., Ferretti, G., Klebanov, I. (1997)
`Creation of Fundamental Strings by Crossing D-branes', {\em
  Phys. Rev. Lett. \bf  79}, 1984--1987, 
hep-th/9705084.

\refce
de Alwis, S.P. (1996) `A note on brane tension and M-theory',
{\em Phys. Lett. \bf B388},  291--295,
hep-th/9607011. 

\refce
de Alwis, S.P. (1997) `Coupling of branes and normalization
 of effective actions in string/M-theory', {\em Phys. Rev. \bf D56},
 7963-7977, hep-th/9705139. 

\refce
de Azcarraga, J.A., Gauntlett, J.P., Izquierdo, J.M., Townsend,
P.K. (1989) `Topological Extensions of the Supersymmetry Algebra for
Extended Objects', {\em Phys. Rev. Lett. \bf 63}, 2443--2446.

\refce
de Roo, M., Suelmann, H., Wiedemann, A. (1993)
` The Supersymmetric Effective Action of
 the Heterotic String in Ten Dimensions', {\em Nucl. Phys. \bf B405},
 326--366,  hep-th/9210099.

\refce
Deser, S., Gomberoff, A., Henneaux, M., Teitelboim, C. (1997)
`p-Brane Dyons and Electric-magnetic Duality', 
hep-th/9712189.

\refce
Deser, S., Henneaux, M., Schwimmer, A. (1998)
 `p-brane Dyons, theta-terms and Dimensional Reduction',
hep-th/9803106.

\refce
de Wit, B. (1998) `Supermembranes and Super Matrix Theory', Lecture at
the 31-st International Symposium Ahrenshoop on the `Theory of
    Elementary Particles', Buckow, to appear in Fortschritte der
    Physik, hep-th/9802073. 

\refce
de Wit, B., Louis, J. (1998) `Supersymmetry and Dualities in Various
Dimensions', lectures at Cargese 97,
hep-th/9801132. 

\refce
Dienes, K. (1997) `String Theory and the Path
 to Unification: A Review of Recent Developments', {\em Phys. Rep.
   \bf 287}, 447--525, 
hep-th/9602045. 

\refce
Dijkgraaf, R. (1997) `Les Houches Lectures on Fields, Strings and
Duality',  hep-th/9703136.

\refce
Dijkgraaf, R., Verlinde E., Verlinde H. (1998)
`Notes on Matrix and Micro Strings', {\em Nucl. Phys. Proc. Suppl. \bf
  62}, 348--362,  hep-th/9709107. 

\refce
Dine, M., Seiberg, N. (1997)
`Comments on Higher Derivative Operators in Some SUSY Field Theories',
 {\em  Phys. Lett. \bf B409},  239--244,
hep-th/9705057.

\refce
Dine, M., Echols, R., Gray, J. (1998) `Renormalization
 of Higher Derivative Operators in the Matrix Model', 
hep-th/9805007.

\refce
Dirac, P.A.M. (1931) `Quantized Singularities in the Electromagnetic
 Field', {\em Proc. Roy. Soc. London \bf A133}, 60--72.

\refce
Di Vecchia, P. (1998) `Duality in Supersymmetric $N=2,4$ Gauge
Theories', hep-th/9803026.

\refce
Dorn, H., Otto, H-J. (1996) `On T-duality for Open Strings in General
 Abelian and Nonabelian Gauge Field Backgrounds', {\em Phys. Lett. \bf
B381}, 81--88,    
 hep-th/9603186.

\refce
Douglas, M.R. (1995) `Branes within Branes',
 hep-th/9512077. 

\refce
Douglas, M.R., Moore, G. (1996) `D-branes, Quivers, and ALE Instantons',
hep-th/9603167.

\refce
Douglas, M.R. (1996) `Superstring Dualities, Dirichlet Branes and
 the Small Scale Structure of Space', 
in the proceedings of the LXIV Les Houches session on `Quantum
Symmetries', hep-th/9610041. 

\refce
Douglas, M.R., Kabat, D., Pouliot, P., Shenker, S.H. (1997)
`D-branes and Short Distances in String Theory', {\em Nucl. Phys. \bf
  B485}, 85--127,
hep-th/9608024. 

\refce
Douglas, M.R. (1997) `D-branes and Matrix Theory in Curved Space', 
talk at `Strings 97',  hep-th/9707228.

\refce
Douglas, M.R., Kato, A., Ooguri, H. (1997) `D-brane Actions on Kahler
Manifolds',  hep-th/9708012. 

\refce
Duff, M.J., Khuri, R.R., Lu, J.X. (1995) `String Solitons',
{\em Phys. Rep. \bf 259}, 213--326, hep-th/9412184. 

\refce
Duff, M.J., Minasian, R. (1995) `Putting String/String Duality to the
Test', {\em Nucl. Phys. \bf B436}, 507--528, hep-th/9406198.

\refce
Duff, M.J. (1995) `Strong/Weak Coupling Duality from the Dual String',
{\em Nucl. Phys. \bf B442}, 47--63,  hep-th/9501030.

\refce
Duff, M.J., Liu, J.T., Minasian, R. (1995)
 `Eleven Dimensional Origin of String/String Duality: A One Loop
 Test', {\em Nucl. Phys. \bf B452}, 261--282, 
hep-th/9506126.

\refce
Duff, M. (1997)  `Supermembranes', 
 Proceedings of the Workshop on `Gauge Theories, Applied
Supersymmetry and Quantum Gravity', Sevrin A. {\em et al} eds., 
Imperial College Press,
hep-th/9611203 . 

\refce
Eguchi, T., Gilkey, P.B., Hanson, A.J. (1980) `Gravitation, Gauge
Theories and Differential Geometry', {\em Phys. Rep. \bf 66}, 213--393.

\refce
Elitzur, S., Giveon, A., Kutasov, D. (1997a) `Branes and N=1 Duality
in String Theory', {\em Phys. Lett. \bf  B400}, 269--274, hep-th/9702014.

\refce
Elitzur, S., Giveon, A., Kutasov, D., Rabinovici, E., Schwimmer,
A. (1997b) `Brane Dynamics and N=1 Supersymmetric Gauge Theory', {\em 
Nucl. Phys. \bf B505}, 202--250, hep-th/9704104.

\refce
 Ferrara, S.,  Harvey, J.A.,  Strominger, A.,  Vafa C. (1995)
`Second-Quantized Mirror Symmetry', {\em Phys. Lett. \bf B361}, 59--65,
 hep-th/9505162.

\refce
F\" orste, S., Louis, J. (1997) `Duality in String Theory', 
 Proceedings of the Workshop on `Gauge Theories, Applied
Supersymmetry and Quantum Gravity', Sevrin A. {\em et al} eds., 
Imperial College Press, hep-th/9612192. 

\refce
Fradkin, E.S., Tseytlin, A. (1985) `Non-linear Electrodynamics from
Quantized Strings', {\em Phys. Lett. \bf B163}, 123--130.

\refce
Friedan, D., Martinec, E., Shenker, S. (1986) `Conformal Invariance,
 Supersymmetry and String Theory',
{\em Nucl. Phys. \bf B271}, 93--165.

\refce
Gauntlett, J.P. (1997) `Intersecting Branes', Lectures at APCTP Winter
School on `Dualities of Gauge and String Theories', Korea, 
hep-th/9705011.  

\refce
Gauntlett, J.P., Gibbons, G.W., Papadopoulos, G., Townsend,
P.K. (1997)
`Hyper-Kahler Manifolds and Multiply-intersecting Branes',
{\em Nucl. Phys. \bf B500}, 133--162, 
hep-th/9702202.

\refce
Gibbons, G.W., Green, M.B., Perry, M.J. (1996)
`Instantons and Seven-branes in Type IIB Superstring Theory',
{\em Phys. Lett. \bf B370}, 37--44, 
hep-th/9511080. 

\refce
Gibbons, G.W., Horowitz, T., Townsend, P.K. (1995)
`Higher-dimensional Resolution of Dilatonic Black Hole Singularities',
{\em Class.Quant.Grav.  \bf 12}, 297-318, 
hep-th/9410073.

\refce
Gibbons, G.W. (1998) `Born-Infeld Particles and Dirichlet p-branes',
{\em Nucl. Phys. \bf B514}, 603--639,
hep-th/9709027.

\refce
Gimon, E., Polchinski, J. (1996)
 `Consistency Conditions for Orientifolds and D-Manifolds', 
{\em Phys. Rev. \bf D54}, 1667--1676, 
hep-th/9601038. 

\refce
Giveon, A., Porrati, M., Rabinovici, E. (1994) `Target Space Duality
 in String Theory', {\em  Phys. Rep. \bf 244 },  77--202,
 hep-th/9401139.

\refce
Giveon, A., Kutasov, D. (1998) `Brane Dynamics and Gauge Theory',
  hep-th/9802067. 

\refce
Green, M.B., Schwarz, J.H. (1984)
`Anomaly Cancellations in Supersymmetric D=10 Gauge Theories and 
 Superstring Theory', {\em  Phys. Lett. \bf B149}, 117--122.

\refce
Green, M.B., Schwarz, J.H. (1985a)
`Infinity Cancellations in SO(32) Superstring Theories', {\em
  Phys. Lett. \bf B151}, 21--25.

\refce
Green, M.B., Schwarz, J.H. (1985b)
`The Hexagon Gauge Anomaly in Type 1 Superstring Theory', {\em
  Nucl. Phys. \bf B255}, 93--114.

\refce
Green, M.B., Schwarz, J.H., Witten, E. (1987) `Superstring Theory',
Cambridge University Press, two volumes.

\refce
Green, M.B. (1991a) `Space-Time Duality and Dirichlet String Theory',
{\em Phys. Lett. \bf B266}, 325--336.

\refce
Green, M.B. (1991b) `Duality, Strings and Point-like Structure', 
talk at 25th Rencontres de Moriond, Tr{\^a}n Thanh V{\^a}n, J. ed,
editions Frontieres.

\refce
Green, M.B., Gutperle, M. (1996)  `Light-cone Supersymmetry and D-branes',
{\em Nucl. Phys. \bf B476}, 484--514, hep-th/9604091.

\refce
Green, M.B., Hull, C.M., Townsend, P.K. (1996) `D-brane Wess--Zumino
Actions,  T-duality and the Cosmological Constant', {\em Phys. Lett.
\bf B382}, 65--72,  hep-th/9604119. 

\refce
Green, M.B., Harvey, J.A., Moore, G. (1997) 
`I-Brane Inflow and Anomalous Couplings on D-Branes', 
{\em Class.Quant.Grav. \bf 14}, 47--52, 
hep-th/9605033.

\refce
Green, M.B. (1997) `Connections between M-theory and Superstrings',
talk at Cargese 97, 
hep-th/9712195.

\refce
Gross, D., Perry, M. (1983) `Magnetic Monopoles in Kaluza-Klein
Theories',
 {\em Nucl. Phys. \bf B226}, 29--48.

\refce
Gubser, S., Klebanov, I., Polyakov, A. (1998) `Gauge Theory
Correlators from Noncritical String Theory', hep-th/9802109. 

\refce
Gutperle, M. (1997) `Aspects of D-Instantons', talk at Cargese 97, 
hep-th/9712156. 

\refce
G\" uven, R. (1992) `Black p-brane Solutions of D=11 Supergravity
Theory', {\em Phys. Lett. \bf B276}, 49--55.

\refce
Hanany, A.,   Witten, E. (1997) `Type IIB Superstrings,
 BPS Monopoles, and Three-Dimensional Gauge Dynamics', {\em
   Nucl. Phys.
\bf B492}, 152--190, hep-th/9611230.

\refce
Harvey, J.A. (1996) `Magnetic Monopoles, Duality, and Supersymmetry',
hep-th/9603086. 

\refce
Hashimoto, A., Klebanov, I. (1997) `Scattering of Strings from D-branes',
{\em Nucl. Phys. Proc. Suppl. \bf 55B}, 118--133, 
 hep-th/9611214.

\refce
Hashimoto, A., Taylor, W. (1997) `Fluctuation Spectra
 of Tilted and Intersecting D-branes from the Born-Infeld Action', 
{\em Nucl. Phys. \bf B503}, 193--219,
hep-th/9703217. 

\refce
Hassan, S.F., Wadia, S.R. (1997) `Gauge Theory Description
 of D-brane Black Holes: Emergence of the Effective SCFT and
    Hawking Radiation', 
hep-th/9712213. 

\refce
Henneaux, M., Teitelboim, C. (1986) `Quantization of Topological Mass
in the Presence of a Magnetic Field', {\em Phys. Rev. Lett. \bf 56},
689--692.

\refce
Ho\v rava, P. (1989a) `Strings on Worldsheet Orbifolds', {\em
  Nucl. Phys. \bf B327}, 461--484. 

\refce
Ho\v rava, P. (1989b) `Background Duality of Open-string Models', {\em
  Phys. Lett. \bf B231}, 251--257.

\refce
Ho\v rava, P., Witten, E. (1996a)  
`Heterotic and Type I String Dynamics from Eleven Dimensions',
{\em Nucl. Phys. \bf B460}, 506--524, hep-th/9510209. 

\refce
Ho\v rava, P., Witten, E. (1996b) `Eleven-dimensional Supergravity
 on a Manifold with Boundary', {\em Nucl. Phys. \bf B475}, 94--114,
hep-th/9603142.  

\refce
Howe, P.S., Lambert, N.D., West, P.C. (1998) `A New Massive
 Type IIA Supergravity from Compactification', {\em Phys. Lett. \bf
   B416}, 303--308, hep-th/9707139.

 \refce 
Hull, C.M.,  Townsend, P.K. (1995)  `Unity of Superstring Dualities',
 {\em Nucl. Phys. \bf B438}, 109--137, hep-th/9410167. 

\refce
Hussain, F., Iengo, R., N{\'u}{\~n}ez, C., Scrucca, C. (1997)
` Interaction of Moving D-Branes on Orbifolds'
 {\em  Phys. Lett. \bf  B409},  101--108, hep-th/9706186. 

\refce
Intriligator, K., Seiberg, N. (1996) `Lectures on
 Supersymmetric Gauge Theories and Electric-Magnetic Duality', {\em
Nucl. Phys. Proc. Suppl. \bf 45B}, 1--28,
hep-th/9509066. 

\refce
Julia, B.L. (1998) `Dualities in the Classical Supergravity Limits',
talk in Cargese 97, hep-th/9805083.

\refce
Kabat, D., Pouliot, P. (1996) `A Comment on Zero-brane Quantum Mechanics',
{\em Phys. Rev. Lett. \bf 77}, 1004-1007, 
hep-th/9603127. 

\refce
Kachru, S., Vafa, C. (1995) `Exact Results for N=2 Compactifications
 of Heterotic Strings', {\em Nucl. Phys. \bf B450}, 69--89, 
hep-th/9505105.

\refce
Kiritsis, E. (1998) `Introduction to Superstring Theory',
to be published by Leuven University Press,  hep-th/9709062. 

\refce
Kitazawa, Y. (1987) `Effective Lagrangian for the Open Superstring
from a 5-point Function',
{\em Nucl. Phys. \bf B289}, 599--608.

\refce
Kogan, I.I., Mavromatos, N.E., Wheater, J.F. (1996)
`D-Brane Recoil and Logarithmic Operators', {\em Phys. Lett. \bf
  B387}, 483--491, hep-th/9606102.

\refce
Leigh, R.G. (1989) `Dirac-Born-Infeld Action for
 Dirichlet $\sigma$-Models', {\em Mod. Phys. Lett. \bf A4},
 2767--2772.

\refce
Lerche, W. (1988)`Elliptic Index and the Superstring Effective Action', 
{\em Nucl. Phys. \bf B308}, 102--126.

\refce
Lerche, W. (1997) `Introduction to Seiberg-Witten Theory and its
Stringy Origin', {\em Nucl. Phys. Proc. Suppl. \bf 55B}, 83--117,
{\em Fortsch. Phys. \bf 45}, 293--340, 
hep-th/9611190. 

\refce
Li, M. (1996a) `Boundary States of D-Branes and Dy-Strings',
{\em Nucl. Phys. \bf B460}, 351--361, hep-th/9510161. 

\refce
Li, M. (1996b) `Dirichlet Boundary State in Linear Dilaton Background',
{\em Phys. Rev. \bf D54},  1644--1646
hep-th/9512042.

\refce
Lifschytz, G. (1996) `Comparing D-branes to Black-branes',
{\em Phys. Lett. \bf B388},  720--726,
hep-th/9604156.

\refce
Lifschytz, G. (1997) `Probing Bound States of D-branes',
{\em Nucl. Phys. \bf B499},  283--297, 
 hep-th/9610125.

\refce
Lifschytz, G., Mathur, S.D. (1997) `Supersymmetry and
 Membrane Interactions in M(atrix) Theory', {\em Nucl. Phys. \bf
   B507}, 621--644, 
hep-th/9612087.

\refce
Lu, J.X. (1977) `Remarks on M-theory Coupling Constants and M-brane
 Tension Quantizations', 
hep-th/9711014. 

\refce
 L{\"u}scher, M.,  Symanzik, K.,  Weisz, P. (1980)
 `Anomalies of the Free Loop Equation in the WKB Approximation',  {\em
 Nucl. Phys. \bf B173}, 365--396.

\refce
L\" ust, D. (1998) `String Vacua with N=2 Supersymmetry in Four
Dimensions', hep-th/9803072. 

\refce
Madore, J. (1995) `An Introduction to Noncommutative Differential
Geometry and its Physical Applications', Cambridge U. Press.

\refce
Maldacena, J.M. (1996) `Black Holes in String Theory', 
{ Princeton U.  PhD thesis}, hep-th/9607235.

\refce
Maldacena, J.M. (1997) `Branes Probing Black Holes', talk at STRINGS'97,
hep-th/9709099.

\refce
Maldacena, J.M. (1997) ` The Large N Limit of Superconformal Field
 Theories and Supergravity', 
hep-th/9711200.

\refce
Manton, N. (1982) `A Remark on the Scattering of BPS Monopoles', {\em
  Phys. Lett. \bf B110}, 54--56.

\refce
Matusis, A. (1997) `Interaction of non-parallel D1-branes',
hep-th/9707135.

\refce
Milnor, J.W., Stasheff, J.D. (1974) `Characteristic Classes',
Princeton U. Press.

\refce
Morales, J.F., Scrucca, C.A., Serone, M. (1997)
`A Note on Supersymmetric D-brane dynamics', 
hep-th/9709063. 

\refce
Morales, J.F., Scrucca, C.A., Serone, M. (1998)
`Scale Independent Spin Effects in D-brane Dynamics', 
hep-th/9801183. 

\refce
Minasian, R., Moore, G. (1997) `K-theory and Ramond-Ramond Charge ', 
{\em JHEP \bf 11}, 002, hep-th/9710230. 

\refce
Mourad, J. (1998) `Anomalies of the SO(32) Five-brane and their Cancellation',
{\em Nucl. Phys. \bf B512}, 199--208, hep-th/9709012. 

\refce
Nakahara, M. (1990) `Geometry, Topology and Physics', Graduate Student
Series in Physics, Inst. of Phys. Publ.

\refce
Nepomechie, R.I. (1985) `Magnetic Monopoles from Antisymmetric-Tensor
Gauge Fields', {\em Phys. Rev. \bf D31}, 1921--1924.

\refce
Olive, D. (1996) `Exact Electromagnetic Duality', 
{\em Nucl. Phys. Proc. Suppl. \bf  45A}, 88--102, 
 hep-th/9508089. 

\refce
Ooguri, H., Yin, Z. (1996) `TASI Lectures on Perturbative String Theories',
presented  at TASI96, hep-th/9612254.

\refce
Paban, S., Sethi, S., Stern, M. (1998) `Constraints From
 Extended Supersymmetry in Quantum Mechanics', 
hep-th/9805018. 

\refce
Periwal, V., Tafjord, Ø. (1996) `D-brane recoil', {\em Phys. Rev. \bf
  D54}, 3690--3692, 
hep-th/9603156.

\refce
Peskin, M. (1997) `Duality in Supersymmetric Yang-Mills Theory',
lectures at TASI96, 
hep-th/9702094. 

\refce 
Polchinski, J. (1995)  `Dirichlet-Branes and Ramond-Ramond Charges',
 {\em Phys. Rev. Lett. \bf 75}, 4724--4727, hep-th/9510017. 

\refce
Polchinski, J.,  Chaudhuri, S., Johnson, C. (1996)
`Notes on D-Branes', lectures at ITP--Santa Barbara, hep-th/9602052.

\refce
Polchinski, J. (1996)  `TASI Lectures on D-Branes', 
lectures at  TASI 96, hep-th/9611050.

\refce
Polchinski, J., Witten, E. (1996) `Evidence for Heterotic-Type I
Duality', {\em Nucl. Phys. \bf  B460},  525--540, 
 hep-th/9510169. 

\refce
Polchinski, J. (1998) `String Theory', Cambridge U. Press., two volumes.

\refce
Polychronakos, A. (1987) `Topological Mass Quantization and Parity
Violation in 2+1 Dimensional QED',
 {\em Nucl. Phys. \bf B281}, 241--252.

\refce
Porrati, M.,  Rozenberg, A. (1998) `Bound States at Threshold
 in Supersymmetric Quantum Mechanics', {\em  Nucl. Phys. \bf  B515},
184--202, hep-th/9708119.

\refce
Pradisi, G., Sagnotti, A. (1989) `Open String Orbifolds', {\em
Phys. Lett. \bf B216}, 59--67. 

\refce
Polyakov, D. (1996) `RR-Dilaton Interaction in a Type IIB Superstring'
    {\em Nucl. Phys. \bf B468}, 155--162,
hep-th/9512028. 

\refce
Rey, S-J. (1997) `Heterotic M(atrix) Strings and Their Interactions', 
{\em Nucl. Phys. \bf B502}, 170--190,  hep-th/9704158.

\refce
Romans, L. (1986)  `Massive N=2a Supergravity in Ten Dimensions',
{\em Phys. Lett. \bf B169}, 374--380.

\refce
Sagnotti, A. (1988) 
in `Non-Perturbative Quantum Field Theory', eds. Mack {\em et al}, Pergamon
Press, p.521.

\refce
Sagnotti, A. (1997) `Surprises in Open-String Perturbation Theory',
{\em Nucl. Phys. Proc. Suppl. \bf B56}, 332--343, hep-th/9702093. 

\refce
Schmidhuber, C. (1996) `D-brane Actions', {\em Nucl. Phys. \bf B467},
146--158,  hep-th/9601003.

\refce
Schwarz, J.H. (1995) `An SL(2,Z) Multiplet of Type IIB Superstrings',
{\em Phys. Lett \bf B360}, 13--18, erratum ibid {\bf B364}, 252, 
hep-th/9508143 

\refce
Schwarz, J.H. (1997a) ` Lectures on Superstring
and M-theory Dualities', lectures at the ICTP Spring School and TASI 96, 
{\em Nucl. Phys. Proc. Suppl.  \bf 55B}, 1--32,  hep-th/9607201.

\refce
Schwarz, J.H. (1997b) `The Status of String Theory', hep-th/9711029.

\refce
Schwinger, J. (1968) `Sources and Magnetic Charge',
 {\em Phys. Rev. \bf 173}, 1536--1544.

\refce
Sen, A. (1996) `U-duality and Intersecting D-branes', {\em
  Phys. Rev. \bf D53}, 2874--2894, hep-th/9511026. 

\refce
Sen, A. (1998) `An Introduction to Non-perturbative String Theory', in
this volume, hep-th/9802051. 

\refce
Sethi, S., Stern, M. (1998) `D-Brane Bound States Redux',
 {\em  Commun. Math. Phys. \bf  194},  675--705, hep-th/9705046.

\refce
Shenker, S.H. (1995) `Another Length Scale in String Theory?', 
hep-th/9509132. 

\refce
Sorkin, R. (1983) `Kaluza-Klein Monopole',
 {\em Phys. Rev. Lett. \bf 51}, 87--90.

\refce
Stelle, K. (1997) `Lectures on Supergravity p-branes',
 lectures  at the 1996 ICTP Summer School, Trieste,  
hep-th/9701088. 

\refce
Stelle, K. (1998) `BPS Branes in Supergravity', 
lectures  at the 1997 ICTP Summer School, Trieste,
hep-th/9803116.

\refce
Strominger, A. (1995) `Massless Black Holes and Conifolds in String Theory',
{\em Nucl. Phys. \bf B451},  96--108, hep-th/9504090. 

\refce
Strominger, A. (1996) `Open P-Branes', {\em Phys. Lett. \bf B383}, 44--47,
hep-th/9512059.

\refce
Strominger, A., Vafa, C. (1996) `Microscopic
 Origin of the Bekenstein-Hawking Entropy', {\em Phys. Lett. \bf
   B379}, 99--104, 
hep-th/9601029

\refce
Taylor, W. (1998) `Lectures on D-branes, Gauge Theory and M(atrices)',
lectures  at the 1997 ICTP Summer School, Trieste, hep-th/9801182. 

\refce
 Teitelboim, C. (1986a) `Gauge Invariance for Extended Objects',
 {\em Phys. Lett. \bf B167}, 63--68.

\refce
 Teitelboim, C. (1986b) `Monopoles of Higher Rank',
 {\em Phys. Lett. \bf B167}, 69--72.

\refce
 Thorlacius, L. (1998) ` Introduction to D-branes',
{\em Nucl. Phys. Proc. Suppl. \bf 61A},  86--98, hep-th/9708078. 

\refce
 Townsend, P.K. (1995) `The Eleven-dimensional Supermembrane Revisited',
{\em Phys. Lett \bf B350}, 184--187, hep-th/9501068.

\refce
 Townsend, P.K. (1996a) `D-branes from M-branes',
{\em Phys. Lett \bf B373}, 68--75, hep-th/9512062.

\refce
Townsend, P.K. (1996b) `Four Lectures on M-theory', 
lectures at the 1996 ICTP Summer
School, Trieste,
hep-th/9612121.

\refce
Townsend, P.K. (1997) `M-theory from its Superalgebra', 
 lectures at  Cargese 97,  hep-th/9712004. 

\refce
Tseytlin, A. (1996a) `On SO(32) Heterotic -
 Type I Superstring Duality in Ten Dimensions', {\em Phys. Lett. \bf
    B367}, 84--90, hep-th/9510173. 

\refce
Tseytlin, A. (1996b) `Heterotic - Type
 I Superstring Duality and Low-energy Effective Actions', {\em
   Nucl. Phys. \bf B467}, 383--398, hep-th/9512081. 

\refce
Tseytlin, A. (1997) `On Non-abelian Generalisation of
 Born-Infeld Action in String Theory', {\em Nucl. Phys. \bf B501}, 41--52 
hep-th/9701125. 

\refce
Vanhove, P. (1997) `BPS Saturated Amplitudes and Non-perturbative
String Theory', talk at Cargese 97, 
 hep-th/9712079. 

\refce
Vafa, C. (1996) `Instantons on D-branes', {\em Nucl. Phys. \bf B463},
435--442, hep-th/9512078. 

\refce
Vafa, C. (1997) `Lectures on Strings and Dualities',
hep-th/9702201. 

\refce
Veltman, M. (1975) `Quantum Theory of Gravitation',
in  { Methods in Field Theory}, Les Houches XXVIII, 
 Balian R.  {\em et al}  eds., North Holland.

\refce
West, P. (1998) `Introduction to Rigid Supersymmetric Theories', 
lectures at TASI97,  hep-th/9805055. 

\refce
West, P. (1998) `Supergravity, String Duality and Brane Dynamics', in
this volume.

\refce
Witten, E. (1979) `Dyons of Charge $e\theta/2\pi$', {\em Phys. Lett. \bf
B86}, 283--287.

\refce
Witten, E. (1995) `String Theory Dynamics In Various Dimensions',
{\em Nucl. Phys. \bf B443}, 85--126, hep-th/9503124.

\refce
Witten, E. (1996a) `Bound States of Strings and  $p$-Branes',
{\em Nucl. Phys. \bf B460}, 335--350, hep-th/9510135.

\refce
Witten, E. (1996b) `Small Instantons in String Theory', {\em
  Nucl. Phys. \bf B460}, 541--559,  hep-th/9511030.

\refce
Witten, E. (1997a) `On Flux Quantization In M-Theory And The Effective
Action', {\em  J. Geom. Phys. \bf 22}, 1--13, hep-th/9609122. 

\refce
Witten, E. (1997b) `Five-Brane Effective Action In M-Theory', 
{\em J. Geom. Phys. \bf 22}, 103--133,  hep-th/9610234.

\refce
Witten, E. (1998) `Anti De Sitter Space and Holography', 
hep-th/9802150. 

\refce 
Yi, P. (1997) `Witten Index and Threshold Bound States of D-Branes',
{\em Nucl. Phys. \bf  B505}, 307--318, hep-th/9704098.

\refce
Youm, D. (1997) 
`Black Holes and Solitons in String Theory', hep-th/9710046.

\refce
Zwanziger, D.  (1968) `Quantum Field Theory of
 Particles with both Electric and Magnetic Charges',
 {\em Phys. Rev. \bf 176}, 1489--1495.

\end{document}